\documentclass[10pt,journal]{IEEEtran}
\usepackage{booktabs}
\usepackage{bbding}
\usepackage{multirow}
\usepackage{graphicx}
\usepackage{algorithm}
\usepackage{algorithmic}
\usepackage{CJKutf8}
\usepackage{amsthm,amsmath,amssymb}
\usepackage{mathrsfs}
\usepackage{hyperref} 
\usepackage{subfigure}
\usepackage{color}
\usepackage{makecell}
\usepackage{float}
\usepackage[table]{xcolor}
\usepackage{caption}
\usepackage{tikz}
\usetikzlibrary{graphs,graphs.standard,quotes}
\usepackage{pgfplots}
\usetikzlibrary{fit,matrix,positioning,calc}
\usepackage{comment}

\begin{document}
\begin{CJK}{UTF8}{gbsn}

\title{Deep Reinforcement Learning for Job Scheduling and Resource Management in Cloud Computing: An Algorithm-Level Review}

%\title{Deep Reinforcement Learning for Job Scheduling and Resource Management in Cloud Computing: A Comprehensive Review}

\author{Yan Gu, Zhaoze Liu, Shuhong Dai, Cong Liu, Ying Wang, Shen Wang, Georgios Theodoropoulos, Long Cheng

\thanks{Y. Gu, Z. Liu, S. Dai and L. Cheng are with the School of Control and Computer Engineering, North China Electric Power University in Beijing, China. E-mail: lcheng@ncepu.edu.cn}
\thanks{C. Liu is with the School of Computer Science and Technology, Shandong University of Technology, China. E-mail: liucongchina@sdust.edu.cn}
\thanks{Y. Wang is with the Institute of Computing Technology, Chinese Academy of Sciences, Beijing, China. E-mail: wangying2009@ict.ac.cn}
\thanks{S. Wang is with the School of Computer Science, University College Dublin, Ireland, E-mail: shen.wang@ucd.ie}
\thanks{G. Theodoropoulos is with Department of Computer Science and Engineering, Southern University of Science and Technology, Shenzhen, China. E-mail: georgios@sustc.edu.cn}
}

%\thanks{Manuscript received April 19, 2005; revised September 17, 2014.}}

%{Shell \MakeLowercase{\textit{et al.}}: Bare Demo of IEEEtran.cls for Computer Society Journals}

\IEEEtitleabstractindextext{%
\begin{abstract}
%Cloud computing has emerged as a pivotal paradigm which can provide scalable and reliable computing services to accommodate the increasing demands of various applications. Due to the dynamic and diversity of incoming computing tasks and cloud resources, job scheduling and resource management are crucial to ensuring system performance and efficient and optimal service delivery. Traditional approaches, including heuristic and meta-heuristic algorithms, often exhibit limitations in dynamic or complex cloud environments due to their dependence on comprehensive environmental knowledge, which is challenging to obtain and maintain in real-time, particularly under rapidly evolving conditions. Recent advancements in Deep Reinforcement Learning (DRL) have demonstrated considerable promise in addressing these challenges by facilitating adaptive decision-making in uncertain and non-stationary environments. This survey provides a comprehensive review of DRL-based algorithms tailored for job scheduling and resource management, analyzing their methodologies, performance, and practical applicability within cloud computing. Furthermore, we identify main emerging trends and future research directions for DRL-based scheduling and resource management, highlighting areas with substantial potential for further innovation. Through this survey, we aim to provide researchers and practitioners with an integrative perspective on the evolving contributions of DRL to job scheduling and resource management, thereby advancing the development of innovative solutions in this rapidly progressing field.

Cloud computing has revolutionized the provisioning of computing resources, offering scalable, flexible, and on-demand services to meet the diverse requirements of modern applications. At the heart of efficient cloud operations are job scheduling and resource management, which are critical for optimizing system performance and ensuring timely and cost-effective service delivery. However, the dynamic and heterogeneous nature of cloud environments presents significant challenges for these tasks, as workloads and resource availability can fluctuate unpredictably. Traditional approaches, including heuristic and meta-heuristic algorithms, often struggle to adapt to these real-time changes due to their reliance on static models or predefined rules. Deep Reinforcement Learning (DRL) has emerged as a promising solution to these challenges by enabling systems to learn and adapt policies based on continuous observations of the environment, facilitating intelligent and responsive decision-making. This survey provides a comprehensive review of DRL-based algorithms for job scheduling and resource management in cloud computing, analyzing their methodologies, performance metrics, and practical applications. We also highlight emerging trends and future research directions, offering valuable insights into leveraging DRL to advance both job scheduling and resource management in cloud computing.
\end{abstract}

}

\maketitle

\IEEEdisplaynontitleabstractindextext
\IEEEpeerreviewmaketitle

%------------------------------------------------------------------------
\section{Introduction}
\label{sec:introduction}
%------------------------------------------------------------------------

%\IEEEPARstart{C}loud computing provides a flexible platform for delivering computing, networking, software, and intelligent services over the Internet~\cite{duan2022distributed}. By leveraging shared resource pools, cloud platforms such as Google App Engine (GAE) and Amazon Elastic Compute Cloud (EC2) enable enterprises to dynamically lease hardware and software resources on-demand. Through virtualization, cloud computing enables elastic resource allocation on a pay-as-you-go basis~\cite{saxena2021op}, effectively adapting to fluctuating workload demands. This flexibility not only reduces IT infrastructure costs but also streamlines management and maintenance, encouraging widespread adoption of cloud solutions for enterprise applications and services. 

% \IEEEPARstart{C}
\textbf{Cloud Computing.}
Cloud computing has fundamentally reshaped the landscape of modern computing, offering flexible, scalable, and cost-effective solutions for data storage, processing, and management. Unlike traditional computing models, where users rely on local servers or on-premises infrastructure, cloud computing provides a distributed environment where computational resources, including servers, storage, and software, are delivered over the internet~\cite{marinescu2022cloud}. This shift to cloud-based services enables organizations and individuals to access powerful computing resources without the need to invest heavily in physical infrastructure, allowing them to scale their operations up or down based on demand.

Cloud computing is underpinned by a variety of service models, including Infrastructure-as-a-Service (IaaS), Platform-as-a-Service (PaaS), and Software-as-a-Service (SaaS), each offering different levels of abstraction and control. These services support a wide array of applications, from enterprise resource planning (ERP) systems to machine learning platforms~\cite{cheng2019scalable,mao2022differentiate}, and have become critical enablers of innovation across industries such as finance, healthcare, and entertainment. As the demand for cloud services continues to grow, cloud providers face the ongoing challenge of managing an increasingly complex and dynamic environment to ensure high performance, reliability, and efficiency.

As cloud computing expands to incorporate edge computing~\cite{duan2022distributed}, and scales to support an ever-growing number of users and applications, effective job scheduling and resource management become critical for ensuring optimal performance and resource utilization. Job scheduling involves the allocation of tasks or jobs to available computing resources in a manner that maximizes efficiency, minimizes response times, and ensures fairness among users. Resource management, on the other hand, focuses on the allocation and optimization of computational resources such as CPUs, memory, storage, and bandwidth to meet the diverse needs of various applications and workloads~\cite{luo2021resource}.

\textbf{Job Scheduling and Resource Management.}
The dynamic and heterogeneous nature of cloud environments makes job scheduling and resource management particularly challenging. Cloud workloads vary significantly in terms of computational intensity, real-time constraints, and data dependencies, which complicates the scheduling process~\cite{lu2024a2c,cheng2022cost}. Furthermore, resources are often distributed across multiple servers, data centers, and geographic locations, making it difficult to ensure consistent performance and effective load balancing. The demand for resources fluctuates based on factors such as workload characteristics, user behavior, and external conditions, requiring adaptive management mechanisms~\cite{chatterjee2023dynamic,lei2020real,murthy2023resource}. In addition, cloud providers must address critical issues like energy efficiency and fault tolerance~\cite{yan2022energy,liu2020low}. As a result, efficient job scheduling and resource management become crucial not only for maximizing resource utilization but also for minimizing operational costs, improving quality of service (QoS), and ensuring compliance with service level agreements (SLAs). These factors highlight the need for intelligent, adaptive solutions that can handle the inherent complexities of cloud environments.

\begin{table*}[!htbp]
\centering
\caption{Comparison of existing reviews and surveys on job scheduling and resource management}
\label{tab:survey}
    \setlength{\tabcolsep}{3.2mm}{
    \resizebox{\linewidth}{!}{
    \begin{tabular}{ccccccc}
        \toprule
        \multicolumn{1}{c}{Reference} & \multicolumn{1}{c}{Task Scheduling} & \multicolumn{1}{c}{Workflow Scheduling} & \multicolumn{1}{c}{Resource Provisioning} & \multicolumn{1}{c}{Resource Scheduling}  &  \multicolumn{1}{c}{Algorithm-Level} & \multicolumn{1}{c}{Reviewed Method}  \\ 
        \midrule      
        ~\cite{yang2020recent} & - & - & \checkmark & \checkmark & - & Metaheuristic \\
        ~\cite{xu2021survey} & - & - & \checkmark & \checkmark & - & Heuristic and DRL \\
        ~\cite{luo2021resource} & - & - & \checkmark & \checkmark & - & Heuristic and DRL \\
        ~\cite{houssein2021task} & \checkmark & \checkmark & - & - & - &  Metaheuristic \\
        ~\cite{pradhan2022survey} & \checkmark & \checkmark & - & - & - & Metaheuristic \\
        ~\cite{liu2020resource} & - & - & \checkmark & \checkmark & - &  Metaheuristic \\
        ~\cite{jamil2022resource} & \checkmark & \checkmark & - & \checkmark & - &  Metaheuristic \\
        ~\cite{afrin2021resource} & \checkmark & - & \checkmark & \checkmark & - & Heuristic and DRL \\
        ~\cite{zhou2024deep} & \checkmark & - & \checkmark & \checkmark & -  & DRL\\
        ~\cite{jalali2024deep} &  \checkmark &  \checkmark &  \checkmark &  \checkmark & - & DRL \\   
        \rowcolor{blue!25} This survey & \checkmark & \checkmark & \checkmark & \checkmark & \checkmark & DRL \\
        \bottomrule
        \end{tabular}}}
\end{table*}

A multitude of methods have been developed to address job scheduling and resource management in cloud computing environments. For job scheduling, conventional rule-based policies—such as round-robin~\cite{mohialdeen2013comparative}—and algorithms like Min-Min~\cite{chen2013user} have been widely utilized due to their simplicity and ease of implementation. To enhance scheduling efficiency, heuristic algorithms (e.g., heterogeneous earliest finish time~\cite{nooriantalouki2022heuristic}) and meta-heuristic algorithms (e.g., genetic algorithms~\cite{hoseiny2021pga}, whale optimization algorithm~\cite{chen2020woa}) have been explored for their ability to find near-optimal solutions in complex scheduling scenarios. Similar approaches have been applied to resource management, including heuristic algorithms based on bin packing~\cite{pooranianNovelDistributedFogBased2017} and Petri nets~\cite{niResourceAllocationStrategy2017}, as well as meta-heuristic methods like genetic algorithms~\cite{yangMultiObjectiveTaskScheduling2020} and particle swarm optimization~\cite{potuOptimizingResourceScheduling2021}, which have demonstrated potential in optimizing resource allocation strategies. Despite their effectiveness, these heuristic and meta-heuristic algorithms exhibit significant limitations in real-time, dynamic cloud environments. Their reliance on prior knowledge and static optimization models renders them less adaptable to the unpredictable nature of cloud systems, especially when task arrival times and resource demands fluctuate rapidly.

To overcome these challenges, Deep Reinforcement Learning (DRL) has emerged as a robust and adaptive alternative for both job scheduling and resource management. DRL combines reinforcement learning with deep neural networks, enabling systems to learn optimal policies through continuous interaction with the environment. By leveraging historical data and real-time feedback, DRL algorithms can make informed decisions based on the current state of the cloud environment, effectively adapting to changes and uncertainties~\cite{cheng2023deep}. This adaptive learning process allows DRL to address the complexities inherent in cloud computing, such as dynamic workloads, resource heterogeneity, and unpredictable demand patterns. Consequently, DRL facilitates the development of intelligent scheduling and resource allocation strategies that optimize resource utilization, enhance system performance, and improve QoS. %The ability of DRL to handle high-dimensional state spaces and learn from real-time data makes it a promising approach for advancing job scheduling and resource management in cloud environments.

\textbf{Related Reviews.}
In recent years, significant advances in job scheduling and resource management have driven extensive applications in the domain of cloud computing. Building on these advances, many reviews and surveys have summarized various developments in this field. To provide a more comprehensive understanding, we have synthesized the insights from these reviews to present a holistic perspective. As illustrated in Table~\ref{tab:survey}, our reviews offers a superior contribution compared to existing works by providing a more comprehensive and detailed analysis of DRL techniques aimed at addressing the challenges of job scheduling and resource management. Specifically, the works~\cite{yang2020recent, xu2021survey, luo2021resource} focus on resource management in network function virtualization, 5G networks, and edge computing, respectively, while neglecting a thorough investigation of job scheduling. In comparison, the surveys in~\cite{houssein2021task, pradhan2022survey} provide detailed analyses of task and workflow scheduling in cloud environments relying on meta-heuristic algorithms, but they fall short by overlooking resource management considerations. Studies by~\cite{liu2020resource, jamil2022resource} offer an extensive overview of resource management and task scheduling, however they do not consider DRL-based approaches. Additionally, works~\cite{afrin2021resource,zhou2024deep, jalali2024deep} review DRL-based methods for resource management and task scheduling in cloud computing. Despite their contributions, these works fall short of providing an algorithm-level review of DRL methods, hindering a thorough understanding of DRL advancements in task scheduling and resource management. Our work bridges this gap by systematically analyzing DRL in job scheduling and resource management from an algorithm-level perspective, providing a structured examination of advancements and methodologies.

\begin{figure*}[h]
\centering
\includegraphics[width=0.8\linewidth]{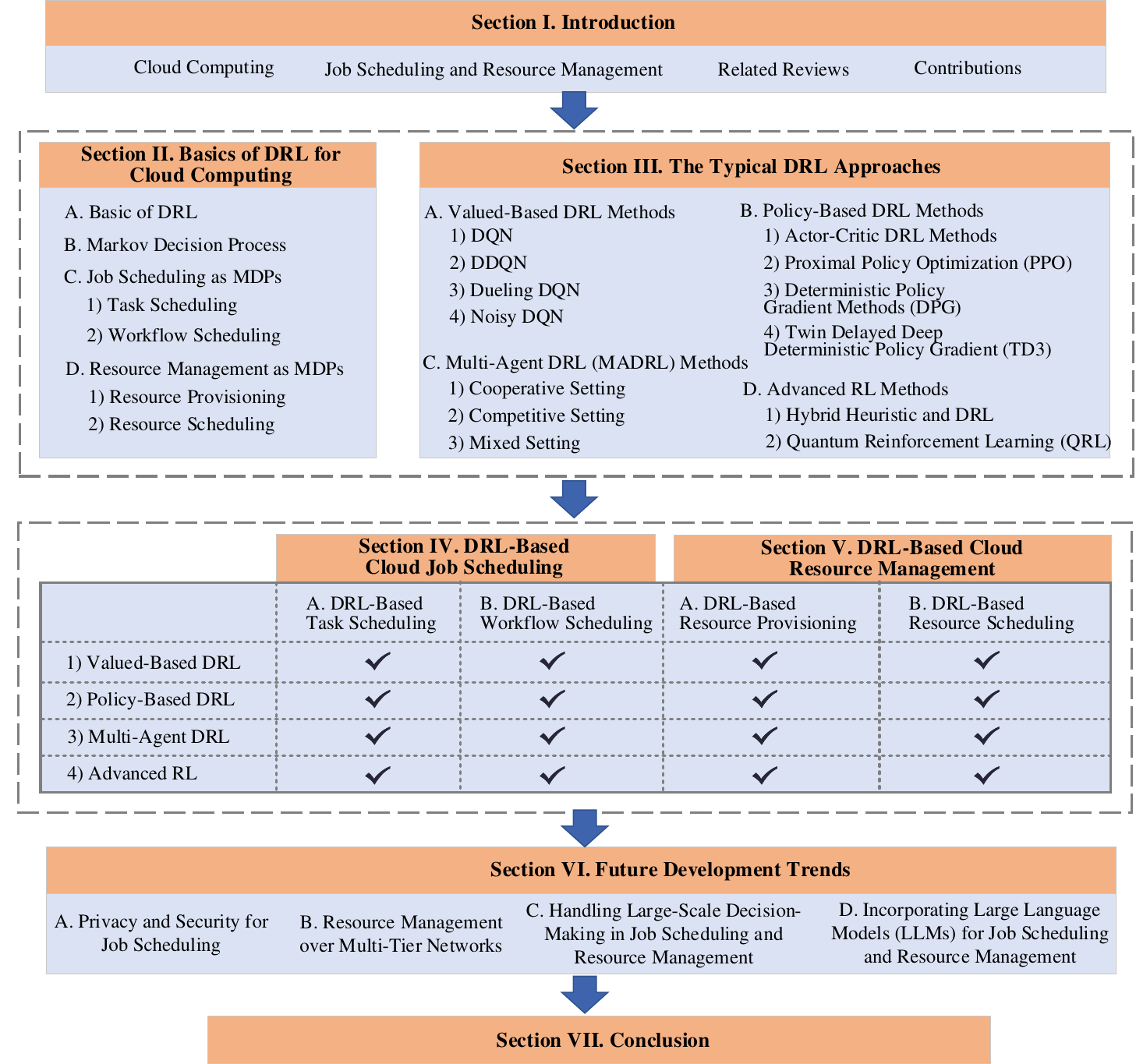}
\caption{The general architecture of this review}
\label{fig:structure}
\end{figure*}

\textbf{Our Contributions.}
In this review, we present a comprehensive analysis of the applications of DRL in job scheduling and resource management, emphasizing the categorization and design principles of various DRL methods. Furthermore, as many studies in edge-cloud computing and edge computing share similar settings with cloud environments, their methods are often applicable to cloud computing scenarios. Consequently, unless explicitly specified, we broaden the scope of this review to include such works. The structure of this review is illustrated in Fig.~\ref{fig:structure}, and our main contributions are summarized as follows:

\begin{itemize}
\item \textit{Algorithm-Level Review and Analysis}: This review provides an in-depth analysis of DRL algorithms, focusing on their applications in job scheduling and resource management within cloud computing. We categorize these algorithms into four main types: value-based methods, policy-based methods, multi-agent approaches, and advanced DRL techniques. This classification not only highlights the unique strengths and capabilities of each approach but also provides a structured framework to tackle the complex challenges inherent in job scheduling and resource management.

\item \textit{Comprehensive Survey of DRL-based Approaches}: This work provides a detailed review of DRL approaches applied to job scheduling and resource management, including task and workflow scheduling, as well as resource provisioning and allocation strategies. We analyze how DRL techniques are customized to optimize these processes, offering insights into their performance, scalability, and adaptability. Furthermore, we examine the application of DRL in cloud and edge computing environments, highlighting the distinct challenges, opportunities, and considerations specific to each context.

\item \textit{Insights into Future Development Trends}: This review highlights critical challenges and opportunities for advancing deep reinforcement learning in job scheduling and resource management within cloud computing. It outlines potential future directions, such as strengthening privacy and security measures, enhancing the robustness and scalability of DRL frameworks, expanding their applicability to dynamic and heterogeneous environments, and improving model interpretability to support practical implementation and real-world adoption.
\end{itemize}

The remainder of this paper is structured as follows: Section~\ref{sec:basic} provides an introduction to the fundamentals of deep reinforcement learning and its relevance to cloud computing. Section~\ref{sec:RL} offers a comprehensive overview of typical DRL algorithms. Section~\ref{job_scheudling} reviews DRL-based methodologies for job scheduling, focusing on both task and workflow scheduling. Section~\ref{resource_management} examines resource management techniques utilizing DRL, including resource provisioning and scheduling strategies. Section~\ref{sec:trends} discusses future directions for advancing DRL in job scheduling and resource management. Finally, Section~\ref{sec:con} provides the conclusion of this review.

%------------------------------------------------------------------------
\section{Basics of DRL for Cloud Computing}
\label{sec:basic}
%------------------------------------------------------------------------
In this section, we introduce the fundamentals of DRL and Markov Decision Processes (MDPs), and provide a general introduction to modeling job scheduling and resource management as MDPs, laying the groundwork for applying DRL in cloud computing.

\subsection{Basics of DRL}\label{2.1}
Reinforcement Learning (RL) has emerged as a powerful paradigm for solving sequential decision-making problems, where an agent learns optimal behaviors through interactions with an environment~\cite{sutton2018reinforcement}. In the RL framework, the agent observes the current state of the environment, selects an action and subsequently receives feedback in the form of a reward which guides future actions. The goal of RL is to develop a policy that maximizes the cumulative reward over time. However, traditional RL methods struggle with high-dimensional state spaces and complex decision problems, especially when state representations and decision-making processes are not straightforward~\cite{li2017deep}.

To address these challenges, DRL combines the principles of RL with the powerful function approximation capabilities of deep neural networks~\cite{ladosz2022exploration}. By using neural networks to approximate value functions or directly model policies, DRL can address problems involving large state and action spaces, commonly found in real-world applications like robotics~\cite{jiang2023learning}, transportation systems~\cite{dai2024MARP}, network control~\cite{liu2021deep2}, and also cloud and edge computing~\cite{liu2021deep}. %The agent in DRL learns from both immediate feedback and delayed rewards which enables it to optimize long-term performance.

% In reinforcement learning, decision-making problems are modeled using a Markov Decision Process (MDP), a framework that captures the sequential interactions between an agent and its environment~\cite{white1989markov}. An MDP is defined by the tuple \( \langle S, A, P, R, \gamma \rangle \), where \( S \) represents the state space, \( A \) the set of actions, \( P(s'|s, a) \) the transition probability from state \( s \) to \( s' \) after taking action \( a \), and \( R(s, a) \) the reward function. The discount factor \( \gamma \in [0, 1] \) balances immediate and future rewards. At each time step \( t \), the agent observes state \( s_t \), takes action \( a_t \), and receives reward \( r_t = R(s_t, a_t) \). The environment then transitions to the next state \( s_{t+1} \) according to \( P(s_{t+1}|s_t, a_t) \). This repeated interaction forms a trajectory \( \tau = (s_0, a_0, r_0, s_1, a_1, r_1, \dots, s_T, a_T, r_T)\) where \( T \) is the length of the episode. The objective of the agent is to maximize the long-term cumulative reward \( G_t \), defined as the sum of discounted future rewards:
% \begin{equation}
%     G_t = \sum_{k=0}^{\infty} \gamma^k r_{t+k},
% \end{equation}
% where \( \gamma \) controls the preference between short-term and long-term rewards. A smaller \( \gamma \) favors immediate gains while a larger \( \gamma \) prioritizes future rewards. The MDP framework provides a foundation for DRL, allowing agents to iteratively improve their policies \( \pi(a|s) \) by maximizing the expected return \( G_t \), leading to better decision-making over time.

\subsection{Markov Decision Process} % tab

In RL, a learning agent interacts with an environment to address sequential decision-making problems. Fully observable environments are typically modeled as MDPs, which are formally defined by a quintuple $(\mathcal{S}, \mathcal{A}, T, \mathcal{R}, \gamma)$. Here, $\mathcal{S}$ denotes the state space, encompassing all possible states the system can occupy, while $\mathcal{A}$ represents the action space, containing all feasible actions an agent can take in any given state. The transition probability function $T$ defines the likelihood of transitioning from one state to another, expressed as $P(s_{t+1} | s_t, a_t)$, given an action $a_t$. The reward function $\mathcal{R}$ assigns a scalar reward $R(s_t, a_t)$ based on the agent's action in a specific state, reflecting the immediate value of that action. Finally, the discount factor $\gamma$, where $0 \leq \gamma \leq 1$, governs the trade-off between immediate and future rewards, with a higher $\gamma$ emphasizing long-term gains and a lower $\gamma$ focusing on short-term outcomes.

At each discrete time step $t$, the agent observes the current state $s_t$, selects an action $a_t$ according to a policy $\pi: \mathcal{S} \rightarrow \mathcal{A}$, and receives a reward $r_t = R(s_t, a_t)$. The environment then transitions to a new state $s_{t+1}$ based on the transition probability $P(s_{t+1} | s_t, a_t)$. The goal of agent is to identify an optimal policy $\pi^*$ that maximizes the expected cumulative reward over time, mathematically expressed as:
\begin{equation}
    \sum_{t=0}^{\infty} \gamma^t r_t(s_t, a_t)
\end{equation}
where $\gamma$ determines the relative importance of future rewards in decision-making. The MDP framework provides a foundation for DRL, allowing agents to iteratively improve their policies \( \pi(a|s) \) by maximizing the expected cumulative reward, leading to better decision-making over time.

% where the discount factor $\gamma$ determines the relative importance of future rewards in decision-making. A higher value of $\gamma$ places greater emphasis on future rewards, leading to actions that are more influenced by long-term outcomes. Conversely, a lower value of $\gamma$ prioritizes immediate rewards, resulting in actions that are more focused on short-term benefits.

\subsection{Job Scheduling as MDPs} 
Job scheduling can generally be abstracted into two levels based on the structure of jobs: task scheduling and workflow scheduling. Workflow scheduling can be considered an extension of task scheduling, as it involves managing more complex dependencies between tasks~\cite{smanchat2015taxonomies}. In dynamic environments, such as those with rapidly fluctuating job arrival rates, scheduling decisions depend solely on the current system state, with future states determined by immediate scheduling actions. This property makes these scenarios particularly well-suited for modeling as MDPs.

\subsubsection{Task Scheduling}
Task scheduling in cloud computing focuses on the allocation of individual tasks to available resources, ensuring that tasks are executed efficiently.

\textbf{Action Space}: Given the sequential arrival of individual tasks over time, the action space at each Markov decision step is designed to schedule the tasks that have arrived at the current moment. Specifically, the scheduler selects an action that determines how the newly arrived task is assigned to available resources or queued for later execution~\cite{liu2024integrated}. The action space $\mathcal{A}$ can be mathematically defined as:
%\begin{equation}
%\mathcal{A} = \{(t_j, v_i) \mid t \in T, v_i \in V\} \cup \{a_q\},
%\end{equation}
\begin{equation}
\mathcal{A} = \{ a \mid a = \text{assign } t \text{ to } v,\; v \in V \}
\end{equation}
Here, \( t \) represents the task to be scheduled at the current decision step, \( v \) is a computational resource to which the task can be assigned, and \( V \) denotes the set of all available computational resources in the system. In some scenarios, the action space $\mathcal{A}$ may vary depending on task-specific constraints. For instance, due to privacy and security considerations, certain tasks may only be assigned to resources in a private cloud~\cite{he2024job}. This dynamic and constraint-aware definition of the action space ensures that scheduling decisions remain feasible and aligned with the specific requirements of tasks and resources.

\textbf{State Space}: 
The task action space is inherently influenced by the state space \( \mathcal{S} \)  of the cloud environment due to the Markov property. Broadly,  \( \mathcal{S} \) comprises two main components: task status \( S_{\text{t}} \) and the states of computational resources \( S_v \). This can be expressed as:

\begin{equation}
\mathcal{S} = \{ S_{\text{t}}, S_v\}
\end{equation}
The task status \( S_{\text{t}} \) captures critical details about each task, such as CPU, memory, and storage requirements, as well as its execution status (e.g., time remaining for completion). In certain computational scenarios, \( S_{\text{t}} \) can be extended to include additional constraints, such as security and privacy requirements or geographical location preferences~\cite{song2005security, chen2018scheduling}. For computational resource states,  \( S_v\) represents the status of available resources, including current availability, expected completion times, computation and storage costs, and other relevant metrics~\cite{mazrekaj2016pricing}.

\textbf{Reward Function}:  In cloud systems, task scheduling often involves multiple optimization objectives. The overall reward function is typically defined as: $\mathcal{R} = \mathcal{F}(o_1,o_2, \ldots)$, where \( o_i \) 
represents an individual optimization objective. Maximization objectives commonly include factors such as memory utilization, storage utilization, and network bandwidth usage. Conversely, minimization objectives focus on metrics like operational costs, task response time, and energy consumption~\cite{guo2005qos, li2024approximate}. In practice, the reward function is often designed as a weighted combination of these objectives, enabling a balanced trade-off between competing goals~\cite{khaleel2024energy}.

\subsubsection{Workflow Scheduling}
Cloud workflows are commonly modeled as Directed Acyclic Graphs (DAGs) and are typically decomposed into subworkflows or subtasks~\cite{deng2015data} during the scheduling process. This decomposition facilitates the parallel execution of workflows across diverse resources in the cloud environment, thereby enhancing resource utilization and workflow execution efficiency.

\textbf{Action Space}: Similar to task scheduling, the action space \(\mathcal{A}\) in workflow scheduling can be abstracted at a high level as:
\begin{equation}
\mathcal{A} = \{ a \mid a = \text{assign } w \text{ to } v,\; v \in V \}
\end{equation}
where $w$ represents the workflow to be scheduled at the current decision step. Generally, the action space in workflow scheduling is typically structured into two decision levels. The first level involves decomposing the workflow \(w\) into subworkflows or tasks \(t\), which can be performed using DRL or other ruled-based algorithms~\cite{topcuoglu2002performance}. The second level focuses on allocating the decomposed subworkflows or tasks to appropriate resources, taking into account task dependencies within the workflow.

\textbf{State Space}: Unlike task scheduling, which primarily focuses on individual tasks, workflow scheduling involves managing entire workflows efficiently by coordinating the execution of decomposed tasks. As a result, the state space for workflow scheduling is inherently more complex and extends beyond that of task scheduling. The state space \( \mathcal{S} \)  for workflow scheduling can be represented as:
\begin{equation}
\mathcal{S} = \{ S_{\text{t}}, S_{\text{w}}, S_v\}
\end{equation}
where \( S_{\text{t}} \) represents the critical details of the decomposed subworkflows or tasks, and  \( S_{\text{w}} \) encapsulates the state of the workflows.

\textbf{Reward Function}: In workflow scheduling, the reward function is tied to various optimization objectives, with makespan and cost often being primary considerations. Unlike traditional task scheduling, workflow scheduling introduces unique challenges, particularly in managing costs. In addition to computation and storage costs, communication costs play a critical role due to the data dependencies between tasks executed across distributed resources. These communication costs, which include data transfer time and bandwidth utilization, can significantly influence workflow performance, especially in geographically dispersed environments~\cite{wu2022communication}.

\subsection{Resource Management as MDPs}
Resource management in cloud computing encompasses two primary functions: resource provisioning and resource scheduling. Resource provisioning involves allocating virtualized resources to accommodate varying workloads and user demands, while resource scheduling focuses on assigning these allocated resources to specific tasks or applications. In dynamic and uncertain cloud environments, resource management decisions are guided by the current state of the system. The evolution of future states depends directly on the actions taken in the present, without reliance on past states, making resource management suitable for representation using MDPs.

\subsubsection{Resource Provisioning}

Through elastic scaling of computational resources, resource provisioning enables cloud systems to adapt to fluctuating workload demands, thereby enhancing overall performance and efficiency.

\textbf{Action Space:} In resource provisioning, the action space depends on the type of scaling. For horizontal scaling, the resource $v$ typically refers to virtual machines or containers, while for vertical scaling, the resource $v$ usually represents CPU, memory, or storage capacity~\cite{mampage2023deep}. Generally, the action space $\mathcal{A}$ be expressed as:
\begin{equation}
    \mathcal{A} = \{a \mid \text{scale up/down/keep unchanged for } v, v \in V\}
\end{equation}
The actions are often represented as discrete integer values, indicating the number of resources to be scaled up or down. 

\textbf{State Space:}
To model the resource provisioning problem, the state space $\mathcal{S}$ can be generally expressed as:
\begin{equation}
    \mathcal{S} = \{S_u, S_n, S_p\}
\end{equation}
where $S_u$ represents the current utilization of system resources, such as CPU, memory, storage, and network bandwidth. $S_n$ denotes the number of virtual machine or container instances running on each server. Additionally, $S_p$ captures performance metrics of the system, including throughput, response time, and energy consumption. 

\textbf{Reward Function:}
The reward function in resource provisioning  incorporates multiple optimization objectives include maximizing resource utilization and system throughput, minimizing infrastructure costs and energy consumption, and maintaining load balance to prevent single-point overloads. Each metric is carefully quantified and integrated into the reward function, ensuring a balanced trade-off between performance, cost efficiency, and system stability.

\subsubsection{Resource Scheduling}
Resource scheduling aims to allocate resources to jobs efficiently, optimizing objectives such as maximizing resource utilization and system throughput.

\textbf{Action Space:}
Give the available resource $V$ and an incoming task $t$ requiring resources at the current step, the action space $\mathcal{A}$ can be defined as:
\begin{equation}
    \mathcal{A} = \{a \mid a = \text{allocate } v \text{ to } t, v \in V\}
\end{equation}
TThe type of resources varies depending on the specific computing scenario. In cloud environments, available resources typically include virtual machines~\cite{liuHierarchicalFrameworkCloud2017}, whereas in edge cloud computing scenarios, the resources could consist of edge devices such as mobile phones~\cite{keDeepReinforcementLearningbased2021} and vehicles~\cite{zhangComputingResourceAllocation2021}. Additionally, resources can be subdivided into more granular components, such as processing units (e.g., CPUs, GPUs), storage, memory, and network bandwidth~\cite{narantuyaMultiAgentDeepReinforcement2022,keMultiAgentDeepReinforcement2022}. 

\textbf{State Space:}  
Similar to task scheduling, resource scheduling involves both the state of the resources \( S_v \) and the state of the tasks \( S_t \). The state space \( \mathcal{S} \) can be represented as:
\begin{equation}
    \mathcal{S} = \{S_v, S_t\}
\end{equation}
where \( S_v \) captures the current status of resources, and \( S_t \) represents the characteristics and requirements of the tasks in the system.

\textbf{Reward Function:}
The design of the reward function for resource scheduling typically considers task requirements and overall system performance. From the task perspective, the primary objectives are to meet QoS and SLA requirements while minimizing the makespan. On the system side, the focus shifts to maximizing resource utilization and minimizing energy consumption and operational costs. 

%------------------------------------------------------------------------
\section{The Typical DRL Approaches}
\label{sec:RL}
%------------------------------------------------------------------------

\begin{figure*}[h]
\centering
\includegraphics[width=1
\linewidth]{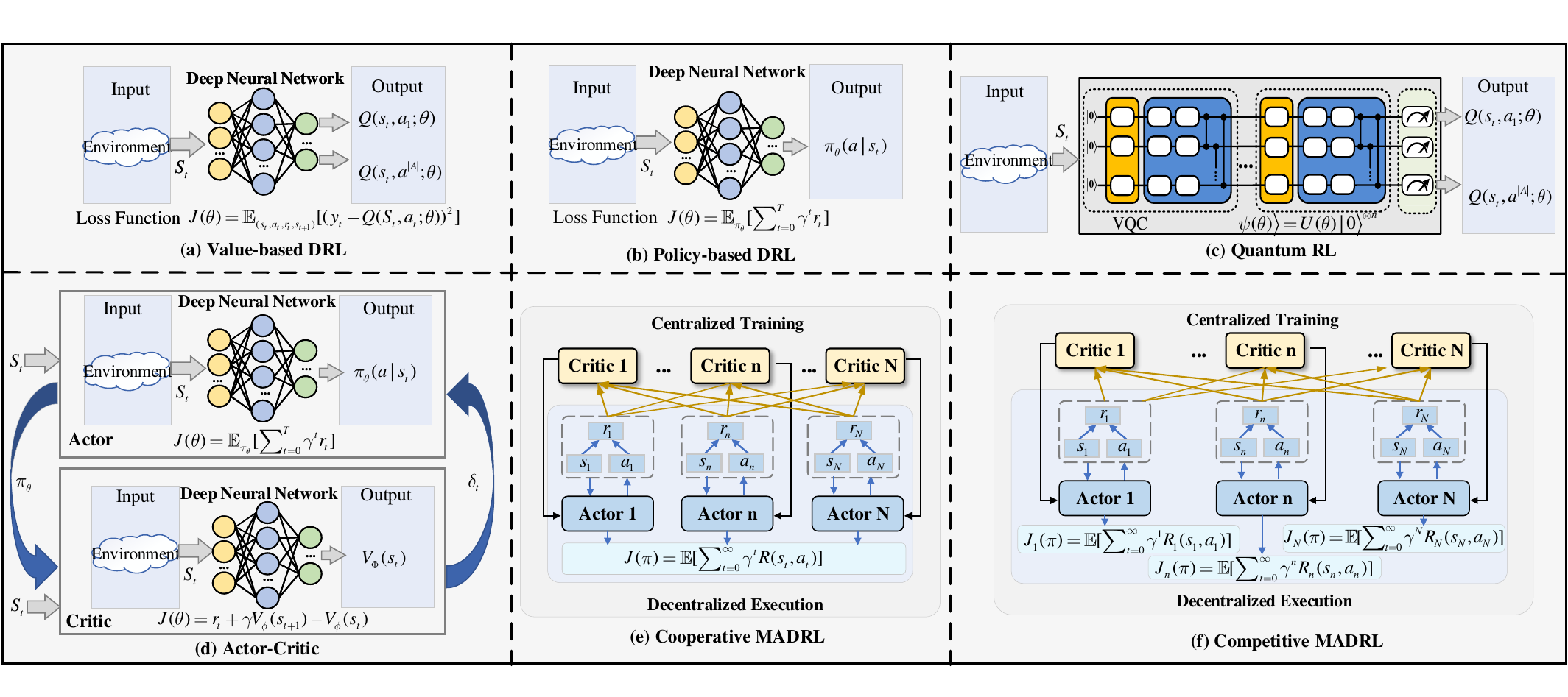}
\caption{Detailed architectures of typical DRL approaches}
\label{fig:drl_architecture}
\end{figure*}

In this section, as illustrated in Fig.~\ref{fig:drl_architecture}, we provide a detailed overview of typical DRL algorithms applied in cloud computing, establishing a foundation for our algorithm-level review on their use in job scheduling and resource management.

\subsection{Valued-Based DRL Methods}\label{2.2}
Value-based DRL methods center around learning a value function which estimates the expected cumulative reward of a state-action pair. As shown in Fig.~\ref{fig:drl_architecture} (a), these methods typically aim to optimize a policy indirectly by approximating the optimal action-value function \( Q^*(s, a) \), which predicts the total expected reward starting from state \( s \), taking action \( a \), and following an optimal policy thereafter. The most widely used value-based method is Deep Q-Network (DQN), which builds upon traditional Q-learning by employing deep neural networks as function approximators~\cite{lan2024rcsearcher}. Several variations of DQN have been proposed to address its limitations, including Double DQN (DDQN), Dueling DQN, and Noisy DQN.

\subsubsection{DQN}
DQN extends Q-learning by utilizing a deep neural network to approximate the Q-function \( Q(s, a; \theta) \) where \( \theta \) represents the parameters of the network. The agent interacts with the environment, storing experiences \( (s_t, a_t, r_t, s_{t+1}) \) in a replay buffer. To update the network, mini-batches of these experiences are sampled and the Q-network is trained to minimize the loss:
\begin{equation}
\label{eq: loss}
    L(\theta) = \mathbb{E}_{(s_t, a_t, r_t, s_{t+1})} \left[ \left( y_t - Q(s_t, a_t; \theta) \right)^2 \right]
\end{equation}
where the target \( y_t \) is calculated as:
\begin{equation}
    y_t = r_t + \gamma \max_{a'} Q(s_{t+1}, a'; \theta^{-})
\end{equation}
and \( \theta^{-} \) refers to the parameters of a target network updated periodically for stability. At each step, the agent selects the action \( a_t = \arg\max_a Q(s_t, a; \theta) \) which maximizes the predicted Q-value for the current state. However, since the same network is used both to select and evaluate actions, this can result in overly optimistic estimates of action values and ultimately lead to overestimation.

\subsubsection{DDQN}
DDQN addresses the overestimation issue in DQN by decoupling the action selection and evaluation processes. In this method, the online Q-network selects the action, while the target network evaluates it to reduce overestimation bias~\cite{chen2024iot}. The target for the loss function is updated as:
\begin{equation}
y_t^{\text{Double}} = r_t + \gamma Q(s_{t+1}, \arg\max_{a'} Q(s_{t+1}, a'; \theta); \theta^{-})
\end{equation}
where the online network selects \( \arg\max_{a'} Q(s_{t+1}, a'; \theta) \), and the target network \( \theta^{-} \) evaluates the Q-value. This decoupling leads to more accurate value estimates and better overall policy learning, particularly in noisy environments.

\subsubsection{Dueling DQN}
In many environments, estimating the value of each action can be inefficient. To address this, the dueling architecture decomposes the Q-value into the state value \( V(s) \) and the action advantage \( A(s, a) \). The Q-value is then expressed as:
\begin{equation}
Q(s, a) = V(s) + A(s, a)
\end{equation}

To ensure the Q-values are uniquely determined, the mean advantage across all actions is subtracted:
\begin{equation}
Q(s, a) = V(s) + \left( A(s, a) - \frac{1}{|\mathcal{A}|} \sum_{a'} A(s, a') \right)
\end{equation}

This structure allows the agent to focus on learning state values, which improves efficiency in environments where the choice of action has less impact. By decoupling state and action evaluation, Dueling DQN enables faster and more stable learning particularly in large action spaces.

\subsubsection{Noisy DQN}
Efficient exploration is essential in DRL, but traditional approaches like \( \epsilon \)-greedy exploration often fail in environments with complex or sparse reward structures. Noisy DQN introduces learnable noise directly into the parameters of the Q-networks to enhance exploration, eliminating the need for manually tuned exploration schedules and enabling more effective exploration throughout training.

In Noisy DQN, the network weights \( W \) are perturbed by adding parameterized noise expressed as:
\begin{equation}
W = \mu + \sigma \cdot \epsilon
\end{equation}
where \( \mu \) and \( \sigma \) are learnable parameters and \( \epsilon \) is drawn from a noise distribution such as Gaussian or factorized noise. This approach ensures that the Q-value estimates vary due to stochasticity in the network and promotes exploration throughout training.

Noisy DQN modifies the standard DQN loss function as described in Equation \ref{eq: loss}, by introducing noisy weights that enhance exploration through added variability in Q-value estimates without the need for manual tuning. As training progresses, the noise adjusts automatically and enables a smooth shift from exploration to exploitation. This approach has shown better performance than standard DQN particularly in tasks with sparse rewards or large action spaces due to its sustained exploration.

\subsection{Policy-Based DRL Methods}\label{2.3}
As illustrated in Fig.~\ref{fig:drl_architecture} (b), policy-based DRL methods focus on directly learning a policy \( \pi(a|s) \) which specifies the probability of taking action \( a \) given state \( s \). Unlike value-based methods that derive a policy indirectly by learning value functions, policy-based methods optimize the policy itself by maximizing the expected cumulative reward over time~\cite{liu2021policy}. These methods are particularly effective in continuous or high-dimensional action spaces, where discretizing the action space for value-based approaches becomes computationally infeasible.

\subsubsection{Actor-Critic DRL Methods}
One of the most widely adopted frameworks within policy-based methods is the Actor-Critic (AC) architecture~\cite{konda1999actor}. In this approach, the actor is responsible for learning the policy \( \pi_\theta(a|s) \), while the critic estimates a value function \( V_\phi(s) \) or the action-value function \( Q_\phi(s, a) \), as illustrated in Fig.~\ref{fig:drl_architecture} (d). The critic provides feedback to the actor on how good the selected actions are, helping to stabilize the policy learning process. The objective of the actor is to maximize the expected return \( J(\theta) \) defined as:
\begin{equation}
J(\theta) = \mathbb{E}_{\pi_\theta} \left[ \sum_{t=0}^{T} \gamma^t r_t \right]
\end{equation}
where \( \gamma \) is the discount factor and \( r_t \) represents the reward at time step \( t \). The policy of the actor is updated by following the gradient of the expected return computed as:
\begin{equation}
\nabla_\theta J(\theta) = \mathbb{E}_{\pi_\theta} \left[ \nabla_\theta \log \pi_\theta(a|s) \cdot \delta_t \right]
\end{equation}
where \( \delta_t = r_t + \gamma V_\phi(s_{t+1}) - V_\phi(s_t) \) is the temporal difference error. The critic updates the value function \( V_\phi(s) \) by minimizing the TD error, thus improving its estimation over time and guiding the actor's policy updates.

% One of the most widely adopted frameworks within policy-based methods is the Actor-Critic (AC) architecture~\cite{konda1999actor}. In this approach, the actor is responsible for learning the policy \( \pi_\theta(a|s) \), while the critic estimates a value function \( V_\phi(s) \) which provides feedback to the actor on how good the selected actions are. This feedback helps stabilize the policy learning process by reducing the variance associated with policy gradient methods.The objective of the actor is to maximize the expected return \( J(\theta) \) defined as:
% \begin{equation}
% J(\theta) = \mathbb{E}_{\pi_\theta} \left[ \sum_{t=0}^{T} \gamma^t r_t \right],
% \end{equation}
% where \( \gamma \) is the discount factor and \( r_t \) represents the reward at time step \( t \). The policy of the actor is updated by following the gradient of the expected return computed as:
% \begin{equation}
% \nabla_\theta J(\theta) = \mathbb{E}_{\pi_\theta} \left[ \nabla_\theta \log \pi_\theta(a|s) A_\phi(s, a) \right],
% \end{equation}
% where \( A_\phi(s, a) \) is the advantage function which measures the relative value of action \( a \) compared to the average action in state \( s \). The role of the critic is to estimate \( A_\phi(s, a) \) by comparing the action-value \( Q_\phi(s, a) \) with the state value \( V_\phi(s) \).

Advantage Actor-Critic (A2C) is a synchronous variant of the standard AC architecture~\cite{mukhopadhyay2019performance}. In A2C, the advantage function \( A(s, a) = Q(s, a) - V(s) \) is explicitly calculated to reduce the bias introduced by the value estimation and to stabilize the updates. By focusing on the advantage rather than raw Q-values or returns, A2C improves the convergence speed and the stability of the learning process. The policy gradient in A2C is computed as:
\begin{equation}
\nabla_\theta J(\theta) = \mathbb{E}_{\pi_\theta} \left[ \nabla_\theta \log \pi_\theta(a|s) \left( Q(s, a) - V(s) \right) \right]
\end{equation}

A2C allows multiple workers to gather experience simultaneously to ensure that updates from each worker contribute to a stable and consistent policy. Asynchronous Advantage Actor-Critic (A3C) builds on this idea but introduces an asynchronous framework where multiple workers interact with their environments independently. In A3C, each worker runs in its own instance of the environment and collects data independently and updates a shared global model. This setup leads to faster updates and reduces the likelihood of workers converging to suboptimal policies, which is a common issue in synchronous methods. By processing their local experiences and updating the global policy and value networks at different times, workers in A3C promote more robust learning and minimize the risk of overfitting to any single trajectory or environment. This results in faster learning speeds and more stable convergence.

\subsubsection{Proximal Policy Optimization (PPO)}
PPO presents a solution to the challenges inherent in policy gradient methods, particularly the risk of unstable or excessively large policy updates. Traditional approaches to policy optimization frequently experience abrupt policy changes after updates, making the learning process more challenging. PPO counteracts this by limiting how much the updated policy can diverge from the previous one using a trust region constraint. This is done via a surrogate objective function that controls the scale of policy changes. At its core, PPO relies on optimizing a clipped objective function, which ensures that policy updates remain within a specified range and do not destabilize the training process. The objective function in PPO is expressed as:

\small
\begin{equation}
L^{\text{CLIP}}(\theta) = \mathbb{E}_{t} \left[ \min \left( r_t(\theta) A_t, \text{clip}(r_t(\theta), 1 - \epsilon, 1 + \epsilon) A_t \right) \right]
\end{equation}
\normalsize
where \( r_t(\theta) = \frac{\pi_\theta(a_t|s_t)}{\pi_{\theta_{\text{old}}}(a_t|s_t)} \) is the probability ratio between the new and old policies, \( A_t \) is the advantage function and \( \epsilon \) is a small hyperparameter that controls the allowable change to the policy. The clipping function ensures that the policy ratio \( r_t(\theta) \) stays within a bounded range \( [1 - \epsilon, 1 + \epsilon] \), thus preventing large updates that could destabilize training.

PPO can be implemented in two different ways, either as PPO-Clip or PPO-Penalty. The most common form PPO-Clip applies the clipped objective as shown above, while PPO-Penalty uses a penalty term to constrain policy changes by minimizing the KL-divergence between the new and old policies:
\begin{equation}
L^{\text{PEN}}(\theta) = \mathbb{E}_{t} \left[ L(\theta) - \beta \, \text{KL}[\pi_{\theta_{\text{old}}}(\cdot|s_t) || \pi_\theta(\cdot|s_t)] \right]
\end{equation}
where \( \beta \) is a penalty coefficient, and the KL-divergence term \( \text{KL}[\cdot || \cdot] \) ensures that the new policy does not deviate significantly from the old one.

\subsubsection{Deterministic Policy Gradient Methods (DPG)}
DPG methods are designed for continuous action spaces by directly learning a deterministic policy \( \mu_\theta(s) \) which selects a specific action for each state. This approach avoids the variance introduced by sampling from a stochastic policy. The goal of DPG is to maximize the expected cumulative reward, formulated as:
\begin{equation}
    J(\theta) = \mathbb{E}_{s_0 \sim p(s)} \left[ \sum_{t=0}^{T} \gamma^t r(s_t, \mu_\theta(s_t)) \right]
\end{equation}
DPG computes the policy gradient using the deterministic policy and the action-value function \( Q(s, a) \), without requiring action sampling. The policy gradient is expressed as:
\begin{equation}
\label{eq: dpg}
    \nabla_\theta J(\theta) = \mathbb{E}_{s \sim \rho^\mu} \left[ \nabla_\theta \mu_\theta(s) \nabla_a Q^\mu(s, a) \big|_{a = \mu_\theta(s)} \right]
\end{equation}

DPG leverages experience replay and target networks to enhance learning stability. While effective in continuous domains, DPG can be prone to local optima due to insufficient exploration. Extensions like Deep Deterministic Policy Gradient (DDPG) address this limitation by incorporating noise-based exploration and lead to more robust performance.

In DDPG, the actor network learns a deterministic policy \( \mu_\theta(s) \) while the critic network estimates the Q-value function \( Q(s, a) \). To ensure exploration, DDPG adds noise to the policy during training typically through an Ornstein-Uhlenbeck process~\cite{onstein-Uhlenbeckprocess}. The critic network is trained using experience replay and target networks (see Equation~\ref{eq: loss}), while the actor network is updated using the deterministic policy gradient (see Equation~\ref{eq: dpg}). These elements help stabilize learning and reduce the variance introduced by noisy gradients.

\subsubsection{Twin Delayed Deep Deterministic Policy Gradient (TD3)}
While DDPG performs well in continuous control tasks, it may still experience overestimated Q-values and result in suboptimal performance. Overestimation can lead to overly optimistic policy updates and degraded performance especially in complex continuous control tasks. TD3 introduces three main improvements to mitigate these issues: clipped double Q-learning, delayed policy updates, and target policy smoothing.

First, TD3 uses clipped double Q-learning to reduce overestimation. Instead of relying on a single Q-network, TD3 uses two Q-networks and updates the policy using the minimum Q-value between the two networks:
\begin{equation}
y = r + \gamma \min_{i=1,2} Q_{\theta_i'}(s', \mu_{\theta'}(s'))
\end{equation}
where \( Q_{\theta_1}(s, a) \) and \( Q_{\theta_2}(s, a) \) represent the estimates from the two critic networks.

The second improvement is delayed policy updates, which decouples the frequency of updates between the actor and critic networks. In TD3, the actor is updated less frequently than the critics, allowing the value function to stabilize before the policy is adjusted. This ensures that the policy is being updated based on more accurate value estimates, further reducing the risk of unstable updates.

Finally, target policy smoothing adds noise to the target actions used in the Q-value updates to prevent the policy from overfitting to narrow peaks in the value function. By applying small and clipped noise to the target actions, TD3 ensures that the policy remains robust to small errors in the value function approximation.

\subsection{Multi-Agent DRL (MADRL) Methods}  

In cloud environments, agents are not isolated entities but interact with others. MARL extends traditional RL methods to such environments, where agents must learn policies in the presence of other agents whose behaviors also evolve during training~\cite{sarkar2022pantheonrl}~\cite{leroy2024imp}. The dynamic nature adds complexity because agents must consider the strategies of others, which can range from cooperative to adversarial. Depending on the relationships between agents, multi-agent settings are typically classified as cooperative, competitive, or mixed, each requiring specific learning strategies and methods.

\subsubsection{Cooperative Setting}
In the cooperative setting, multiple agents share a common objective and their goal is to optimize a joint reward function, as depicted in Fig.~\ref{fig:drl_architecture} (e). This is typically formulated as a Decentralized Partially Observable Markov Decision Process (Dec-POMDP), where agents operate under partial observability and need to collaborate based on local observations to achieve a global goal~\cite{kraemer2016multi}.

A Dec-POMDP is defined by a tuple \( \langle I, S, \{A_i\}_{i \in I}, P, R, \{O_i\}_{i \in I}, \gamma \rangle \). In this framework, \( I \) denotes the set of agents involved in the decision-making process, while \( S \) represents the set of global states in which the system can exist. Each agent \( i \in I \) is associated with a distinct action set \( A_i \), allowing it to select from a range of possible actions. The joint actions of all agents are denoted by \( \mathbf{a} = (a_1, a_2, \dots, a_N) \), where each \( a_i \in A_i \) corresponds to the action chosen by agent \( i \). The state transition function denoted \( P(s'|s, \mathbf{a}) \), models the probability of transitioning from state \( s \) to a new state \( s' \) given the joint action \( \mathbf{a} \). A shared reward function \( R(s, \mathbf{a}) \) determines the immediate reward based on the current state \( s \) and the joint actions of the agents. Each agent \( i \) receives observations from the environment, with the observation function \( O_i \) mapping the global state and joint actions to an observation specific to the agent. Finally, the discount factor \( \gamma \) balances future rewards against immediate ones, shaping the optimization of agent strategies.

The objective in cooperative MARL is to maximize the cumulative reward across all agents:
\begin{equation}
J(\pi) = \mathbb{E} \left[ \sum_{t=0}^{\infty} \gamma^t R(s_t, \mathbf{a}_t) \right]
\end{equation}
where \( \pi = \{\pi_1, \pi_2, \dots, \pi_N\} \) represents the joint policy of all agents. Given the complexity of coordinating decentralized agents, common techniques in this setting involve centralized training with decentralized execution (CTDE). This framework enables agents to learn jointly using shared information during training, but act independently during execution~\cite{Shen2023MultiUAVCS}.

A prominent example of cooperative MARL is Multi-Agent Deep Deterministic Policy Gradient (MADDPG), which extends DDPG to multi-agent settings by introducing a shared critic that considers the joint actions and states of all agents~\cite{li2019robust}. With the actor decentralized, each agent is capable of learning its own policy autonomously:
\begin{equation}
    \nabla_{\theta_i} J(\theta_i) = \mathbb{E} \left[ \nabla_{\theta_i} \log \pi_{\theta_i}(a_i | o_i) Q_{\theta_i}(\mathbf{s}, \mathbf{a}) \right]
\end{equation}
where \( \mathbf{a} = (a_1, a_2, \dots, a_N) \) and \( Q_{\theta_i} \) is the centralized Q-function. The decentralized nature of the actor allows each agent to independently acquire its policy, which is particularly useful in scenarios such as multi-server or multi-cloud task parallelization.

\subsubsection{Competitive Setting}
In the competitive setting, agents have conflicting objectives, where each agent seeks to maximize its own reward often at the expense of others~\cite{daskalakis2020independent}. The corresponding architecture is presented in Fig.~\ref{fig:drl_architecture} (f). This is commonly modeled as a zero-sum game where the sum of the rewards for all agents at any time step is zero. The formal objective for agent \( i \) in a competitive setting is:
\begin{equation}
\begin{aligned}
J_i(\pi_i) &= \mathbb{E} \left[ \sum_{t=0}^{\infty} \gamma^t r_i(s_t, a_t) \right] \\
\text{subject to} \quad \sum_{i=1}^{N} r_i(s, a) &= 0 \quad \text{for all } s \text{ and } a
\end{aligned}
\end{equation}

Furthermore, in competitive environments characterized by adversarial agent actions, a central focus is the determination of the Nash equilibrium, a set of strategies from which no agent has an incentive to unilaterally deviate~\cite{Ye2023DistributedNE}. Competitive multi-agent DRL uses techniques such as policy gradient methods to optimize agents’ policies in adversarial environments. A notable strategy is Self-Play where agents train by competing against themselves~\cite{hu2020other}. Specifically, the agents continuously update their policies by playing against increasingly skilled versions of themselves, leading to higher-level strategies over time. This approach has been successful in applications such as cloud resource allocation and scheduling, where agents learn complex strategies for efficient task distribution in highly dynamic environments. In general, adversarial training and min-max optimization are frequently employed in competitive MARL settings~\cite{zhang2020model}, where agents aim to maximize their own resource usage or task completion while minimizing the efficiency of competing agents:
\begin{equation}
    \min_{\pi_1} \max_{\pi_2} \mathbb{E} \left[ \sum_{t=0}^{T} \gamma^t R(s_t, \pi_1(s_t), \pi_2(s_t)) \right]
\end{equation}
this competitive framework applies to multi-cloud task scheduling, load balancing, and resource allocation, where agents compete to optimize resource usage, dynamically adapting to workload or network fluctuations.

\subsubsection{Mixed Setting}
In the mixed setting, agents must navigate a balance between cooperation and competition, as their objectives may align with some agents while conflicting with others~\cite{zhang2020td3}. As a result, the reward function is neither purely shared nor purely adversarial, leading to the need for multi-objective optimization. For each agent \( i \), the reward function may combine both individual and joint rewards:
\begin{equation}
    r_i(s, \mathbf{a}) = \alpha_i r_{\text{shared}}(s, \mathbf{a}) + (1 - \alpha_i) r_i(s, a_i)
\end{equation}
where \( \alpha_i \) is a weight balancing the shared and individual rewards. Such scenarios frequently arise in real-world applications, where agents must balance both cooperation and competition depending on the specific task at hand. For example, in resource management, agents may collaborate to maximize overall system efficiency but compete for limited resources. The challenge in mixed settings is to dynamically adjust between cooperation and competition depending on the context, making it a highly challenging and significant area of study in MARL.

\begin{figure*}[!t]
\centering
\includegraphics[width=1\linewidth]{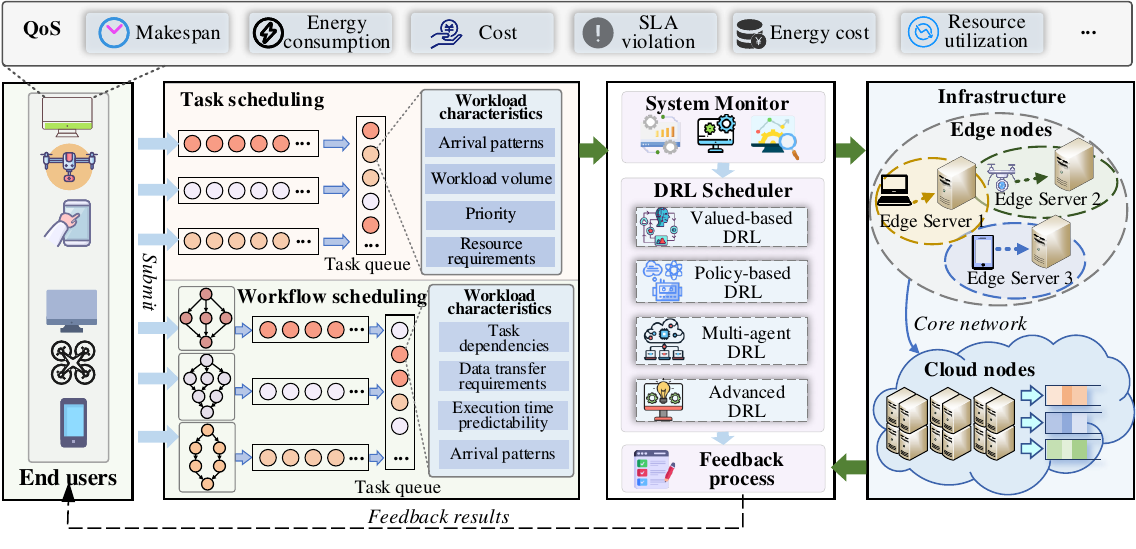}
\caption{The system architecture of job scheduling in cloud computing}
\label{fig:task_scheduling}
\end{figure*}

\subsection{Advanced RL Methods}
\subsubsection{Hybrid Heuristic and DRL}
The integration of heuristic methods with DRL offers a powerful framework for addressing complex optimization problems, leveraging the unique strengths of both approaches. Heuristic algorithms fundamentally search strategies driven by rule-based or experience-informed guidance, excel at swiftly generating approximate solutions within large or high-dimensional search spaces. These algorithms are highly efficient, flexible, and computationally economical, making them well-suited for providing rapid and near-optimal solutions. In contrast, DRL methods excel in learning optimal policies through sequential decision-making and are capable of continuous refinement based on cumulative feedback. The collaboration between heuristic algorithms and DRL can be established through several cooperative paradigms, each contributing to different stages of the problem-solving process.

\textbf{Preliminary Solution Generation and Problem Decomposition: } Before the DRL agent begins its training, heuristic algorithms can provide a initialization by decomposing the task or generating an approximate solution. In scenarios such as cloud workflow scheduling, heuristics can segment tasks based on resource requirements or workflow dependencies, effectively reducing the search space the DRL agent must navigate. This pre-processing phase enables DRL to initiate from a feasible solution space, accelerating the training process and reducing the exploration burden. 

\textbf{Enhanced Action Space Exploration: }During the DRL training phase, heuristic methods can serve as auxiliary tools, guiding the agent’s exploration within the action space. Traditional exploration strategies such as random sampling, often result in inefficient search trajectories particularly in high-dimensional environments or those with sparse rewards. Heuristic algorithms can mitigate these challenges by directing the agent toward regions of the action space with high potential rewards, thus prioritizing valuable paths and enhancing the efficiency of exploration. This heuristic-guided approach not only accelerates training but also increases the likelihood of discovering high-value solutions in complex, reward-scarce environments.

\textbf{Targeted Adjustment and Solution Enhancement: }Following the initial DRL optimization, heuristic algorithms can further refine the solutions generated. While DRL is effective at identifying broadly optimal solutions within complex spaces, it may lack the precision necessary for detailed improvements in specific regions of the solution space. Here, heuristic techniques can be applied to make targeted adjustments, leveraging their adaptability to enhance specific task arrangements or resource allocations. This staged optimization, in which DRL first generates an initial broad solution followed by heuristic-driven local refinements, leads to superior solution quality and greater adaptability to task-specific constraints.

In summary, the hybrid integration of heuristic methods and DRL establishes a layered, adaptable optimization approach, allowing agents to navigate and solve complex, multi-dimensional tasks more effectively. This synergy is especially advantageous in large-scale applications, where it accelerates convergence, enhances solution quality, and provides a versatile framework adaptable to diverse problem landscapes.

\subsubsection{Quantum Reinforcement Learning (QRL)}
With the rapid advancements in quantum technology, the advantages of applying QRL for optimization tasks are increasingly evident. One of the central techniques in QRL is the use of Quantum neural networks (QNNs), which can effectively embed DRL’s state space into a quantum computational framework. Due to the characteristics of quantum circuits, a single state can often be embedded multiple times within the QNN, thus allowing for a richer representation and more diverse exploration. A key feature of QNNs in this context is the quantum variational approach, where quantum variational gates and entanglement gates replace the traditional neurons used in classical neural networks, as shown in Fig.~\ref{fig:drl_architecture} (c). This shift significantly improves the efficiency of trainable parameters, as quantum gates allow for parallel computation and entanglement, enabling more complex state relationships to be represented with fewer resources.

Mathematically, the variational embedding can be expressed as a parameterized quantum circuit \( U(\theta) \) acting on a quantum state \( | \psi \rangle \)~\cite{Morvan2024py}. For example, an initial quantum state can be transformed by a sequence of variational gates:
\begin{equation}
     \psi(\theta) \rangle = U(\theta) | 0 \rangle^{\otimes n}
\end{equation}
where \( \theta \) represents the trainable parameters across the circuit, and \( | 0 \rangle^{\otimes n} \) is the initial quantum state. Through repeated application of entanglement and rotation gates, such as CNOT and parameterized rotation gates \( R(\theta) \), the QNN efficiently learns complex mappings of the DRL state space. Each layer in the QNN introduces entanglement across qubits, thereby capturing correlations within the data that are not easily accessible to classical neural networks.

Compared to traditional DNNs in DRL, QNNs in QRL can achieve comparable or even superior performance with significantly fewer parameters. Variational gates enable this efficiency by leveraging quantum entanglement and superposition, allowing QNNs to capture complex state relationships without the heavy computational load of DNNs. This efficiency makes QRL a promising approach as quantum technology advances~\cite{da2024demonstration}.
%------------------------------------------------------------------------

\section{DRL-based Cloud Job Scheduling}\label{job_scheudling}

\begin{table*}[!t]
\centering
\caption{Summary of typical works on DRL-based task scheduling}
\label{tab:task_scheudling}
    \setlength{\tabcolsep}{3.2mm}{
    \resizebox{\linewidth}{!}{
    \large 
    \begin{tabular}{cccccccc}
        \toprule
        \multicolumn{1}{c}{\multirow{2}{*}{Reference}} & \multicolumn{1}{c}{\multirow{2}{*}{Method}} & \multicolumn{1}{c}{\multirow{2}{*}{Others}} & \multicolumn{1}{c}{\multirow{2}{*}{Environment}} & \multicolumn{4}{c}{Optimization objective} \\
        & & & & Makespan & Cost & Energy consumption & Others \\
        \midrule      
        ~\cite{mangalampalli2024drlbtsa} & DQN & - & Cloud computing & \checkmark & - & \checkmark & SLA violation \\
        ~\cite{cheng2018drl} & DQN & - & Cloud computing & \checkmark & - & - & Energy cost \\
        ~\cite{yan2022energy} & DQN & - & Cloud computing & \checkmark & - & - & Energy cost \\
        ~\cite{kang2021adaptive} & DQN & Online learning & Cloud computing & \checkmark & - & \checkmark & Resource utilization \\
        ~\cite{cheng2023deep} & DQN & - & Cloud computing & \checkmark & \checkmark & - & - \\
        ~\cite{ran2022optimizing} & DQN variants & - & Cloud computing & \checkmark & - & \checkmark & - \\
        ~\cite{oudaa2021agent} & DQN variants & Quantile regression & Cloud computing & \checkmark & - & \checkmark & - \\
        ~\cite{yang2021deep} & DDQN & Greedy optimization & Cloud computing & \checkmark & \checkmark & - & - \\
        ~\cite{tong2021ddqn} & DDQN & - & Cloud computing & \checkmark & - & - & Task completion rate \\
        ~\cite{han2023task} & DQN & - & Edge computing & \checkmark & - & - & Users served by base stations\\
        ~\cite{yuan2021online} & DQN & Online learning & Edge computing & \checkmark & - & - & Avoiding severe task starvation\\
        ~\cite{zeng2023improved} & DDQN & PSO & Edge computing & \checkmark & - & - & - \\
        ~\cite{zhang2018double} & DDQN & - & Edge computing & - & - & \checkmark & Training efficiency \\

        ~\cite{jin2021optimal} & PPO & - & Cloud computing & \checkmark & - & \checkmark & - \\
        ~\cite{yang2022cloud} & PPO & - & Cloud computing & - & - &  \checkmark & - \\
        ~\cite{zhao2021deep} & PPO & - & Cloud computing & \checkmark & \checkmark & - & Renewable energy utilization; deadline violation \\ 
        ~\cite{ran2019slas} & DDPG & - & Cloud computing & \checkmark & - & - & CPU utilization standard deviation \\    
        ~\cite{zhao2022performance} & DDPG & - & Cloud computing & \checkmark & \checkmark & - & - \\
        ~\cite{zhang2024etpam} & SAC & GCN & Edge-Cloud collaboration & \checkmark & - & \checkmark & - \\
        ~\cite{lu2024a2c} & A2C & - & Edge-Cloud collaboration & - & - & - & Server resource utilization; task rejection rate \\
        % ~\cite{lu2024a2c} & A2C & - & Edge-Cloud Collaboration & - & - & - & \begin{tabular}[c]{@{}l@{}}server resource utilization; \\ Task rejection rate\end{tabular} \\
        ~\cite{tuli2020dynamic} & A3C & - & Edge-Cloud collaboration & \checkmark & \checkmark & \checkmark & SLA Violations \\
        % ~\cite{zhao2021deep} & PPO & - & Edge-Cloud Collaboration & \checkmark & \checkmark & - & \begin{tabular}[c]{@{}l@{}}Renewable energy utilization; \\ deadline violation\end{tabular} \\
        ~\cite{zhang2019online} & DDPG & - & Edge computing & - & \checkmark & - & System utility \\  
        ~\cite{chen2022dynamic} & DDPG & - & Edge-Cloud collaboration & - & - & - & Load forward to the cloud server \\
        ~\cite{balla2021reliability} & MADRL & - & Cloud computing & - & - & - & \begin{tabular}[c]{@{}l@{}}Reliability; network communication  \\ overhead; resource utilization\end{tabular} \\
        % ~\cite{balla2021reliability} & MADRL & - & Cloud computing & - & - & - & Reliability; network communication overhead; resource utilization \\ 
        ~\cite{jung2021orchestrated} & MADRL & - & Cloud computing & - & \checkmark & \checkmark & Fairness  \\
        ~\cite{gergely2024multi} & MADRL & - & Cloud computing & \checkmark & - & - & Resource utilization \\
       
        ~\cite{zhang2023multi} & MADRL & - & Edge computing & \checkmark & - & \checkmark & Throughout rate \\ 
        ~\cite{tang2022distributed} & D3RQN & - & Edge cloud & \checkmark & - & \checkmark & Utility \\ 
        ~\cite{xu2021multi} & MADRL & - & Edge computing & - & \checkmark & - & QoS satisfaction rates \\
        ~\cite{zhang2024lsia3cs} & MADRL & SA & Edge-Cloud collaboration & \checkmark & - & \checkmark & Throughput rate \\ 
        ~\cite{li2024load} & MADRL & - & Edge-Cloud collaboration & \checkmark & - & \checkmark & - \\ 
        ~\cite{li2023task} & MAPPO & GNN & Edge computing & \checkmark & - & - & Resource efficiency \\ 
        ~\cite{niu2023multiagent} & Multi-Agent PPO & Meta learning & Edge computing & - & - & - & Resource utilization \\
        ~\cite{gong2021multi} & MADDPG & - & Edge computing & \checkmark & - & \checkmark & - \\ 
        % ~\cite{lim2024adaptive} & MADRL & - & edge cloud & \checkmark &  \checkmark & - & Overall utility \\
        % ~\cite{fellir2020multi} & MADRL & - & cloud-fog computing & - & - & - & Resource utilization \\     
        ~\cite{huang2021deep} & D3QN & Adversarial imitation learning & Cloud computing & \checkmark & - & \checkmark & - \\ 
        % ~\cite{rjoub2019deep} & DQN & LSTM & cloud computing & - & - & - & \begin{tabular}[c]{@{}l@{}}CPU usage cost; \\ RAM memory usage cost\end{tabular} \\ 
        ~\cite{rjoub2019deep} & DQN & LSTM & Cloud computing & - & - & - & CPU usage cost; RAM memory usage cost \\  
        ~\cite{li2023batch} & Distributional RL & Quantile regression & Cloud computing & \checkmark & - & - & Load balancing; success rate \\
        ~\cite{mao2024dl} & Double-level DRL &  Interactive training strategy & Cloud computing & - & \checkmark & - & Computation efficiency\\
        ~\cite{tang2020representation} & DDQN & Representation learning & Edge computing & - & - & \checkmark & SLA \\
        ~\cite{qi2020scalable} & A3C & Multi-task learning & Edge-Cloud collaboration & \checkmark & \checkmark & - & Load imbalance value \\
        
        % ~\cite{zhang2017energy} & PPO & & edge cloud & 1 & 1 & 1 \\
        \bottomrule
        \end{tabular}}}
\end{table*}

In this section, we review existing works on DRL-based task scheduling and workflow scheduling, which form the core of job scheduling in cloud computing. Task scheduling involves assigning individual tasks to available resources to optimize system performance, while workflow scheduling manages interdependent tasks that must execute in a specific order, focusing on task coordination and dependency resolution. As illustrated in Fig.~\ref{fig:task_scheduling}, the scheduling process starts with user requests submitted to the cloud platform. DRL-based approaches allow the scheduler to dynamically analyze workload characteristics and resource availability, enabling efficient task allocation while maintaining adherence to QoS requirements.

\subsection{DRL-Based Task Scheduling}
Through continuous interaction with the environment and learning from experience, DRL has demonstrated exceptional effectiveness in addressing task scheduling challenges in cloud and edge computing environments~\cite{mangalampalli2024drlbtsa, cheng2018drl}. Below, we review key works that employ DRL for task scheduling, with a summarized overview presented in Table~\ref{tab:task_scheudling}.

% Traditional heuristic algorithms, while effective for batch processing, face significant limitations for real-time tasks due to the computational overhead associated with solving complex optimization problems. These limitations are further compounded by the inherently unpredictable nature of cloud environments, where job arrival times and resource demands are unknown in advance. In contrast, DRL excels in such dynamic settings by continuously interacting with the environmenta and learning from experience without relying on prior knowledge. This adaptive capability allows DRL to make informed real-time scheudling decisions, thereby optimizing task scheduling strategies in complex, large-scale, and multi-user environments.
% As a prominent approach grounded in value-function learning, DQN has been widely applied in cloud and edge computing environments. By processing inputs like currently resource utilization and task queue lengths, DQN generates optimal scheduling decisions, determining the most suitable resources for task execution. 

\subsubsection{Valued-Based DRL for Task Scheduling} 

Value-based DRL techniques, such as DQN and its variants, have demonstrated robust decision-making capabilities by estimating state-action value functions for task scheduling. Specifically, these methods focus on learning an optimal policy by approximating the action-value function, which captures the expected reward of selecting specific task scheduling actions, such as assigning tasks to resources or determining execution orders, across various scheduling scenarios~\cite{zhou2024deep}. Extensive studies have investigated the application of valued-based DRL to enhance task scheduling strategies in cloud computing, achieving substantial improvements in task execution efficiency, and resource utilization. As a foundational approach within valued-based DRL, DQN has been effectively utilized to address the inherent challenges of task scheduling including dynamic workload characteristics~\cite{yan2022energy}, workload balancing~\cite{kang2021adaptive}, and the unpredictable nature of cloud environments~\cite{cheng2023deep}. For instance, the work~\cite{mangalampalli2024drlbtsa} successfully applies DQN to minimize energy consumption, makespan and SLA. Likewise, the work~\cite{cheng2018drl} introduces a DQN-based method that employs a two-stage processor, where the first stage assigns tasks to appropriate server clusters and the second stage selects specific servers within those clusters for execution. In the work~\cite{yan2022energy}, a DQN-driven energy-aware scheduling approach is introduced, which dynamically allocates tasks to optimal VMs based on real-time workload characteristics and resource availability, thereby reducing energy consumption while ensuring QoS. To further adapt to dynamic workloads, the work~\cite{kang2021adaptive} develops an adaptive DQN framework, which modifies the discount factor $\gamma$ to optimize energy usage in response to workload fluctuations. Furthermore, the work~\cite{cheng2023deep} presents a DQN-based preemptive method aimed at optimizing job execution cost and response time.

While DQN has demonstrated effectiveness in task scheduling, various DQN variants have also been extensively employed to further improve scheduling quality in cloud environments. These variants have been designed to address specific limitations of conventional DQN approaches, such as Q-value overestimation and the need for adaptation to specialized tasks. For example, the work~\cite{ran2022optimizing} presents a parameterized action space-based DQN (PADQN) framework for jointly optimizing task dispatching and cooling regulation. PADQN specifically addresses the hybrid action space problem by concurrently managing the discrete actions involved in task dispatching and the continuous actions required for cooling regulation. In another work~\cite{oudaa2021agent}, a quantile regression DQN (QR-DQN) network is employed to determine an optimal long-term scheduling strategy, effectively addressing the uncertainties in cloud workload patterns. Given that conventional DQN approaches often suffer from Q-values overestimation, DDQN has been adopted to mitigate this issue. For instance, the work~\cite{yang2021deep} combines DDQN with the greedy optimization strategy for online task scheduling, where DDQN handles task assignments and the greedy algorithm focuses on task execution. Similarly, the work~\cite{tong2021ddqn} applies DDQN to improve task completion rates while simultaneously reducing average response time, demonstrating its effectiveness in handling complex scheduling scenarios under dynamic cloud conditions.

% In dynamic and heterogeneous edge computing environments, task scheduling presents significant challenges due to factors such as resource availability, fluctuating network conditions, energy consumption, and communication delays. DQN-based methods have shown strong decision-making capabilities by adapting to resource variability and efficiently distributing workloads across multiple edge clouds. 

Valued-based DRL methods have also been effectively applied in edge computing environments to address challenges such as resource constraints, fluctuating network conditions, energy consumption, and communication delays. For instance, the work~\cite{han2023task} employs DQN to efficiently schedule tasks to mobile edge computing (MEC) servers, optimizing both delay and bandwidth utilization. Similarly, the work~\cite{yuan2021online} presents a DQN-based method designed to mitigate task delays and alleviate task starvation caused by fluctuating network conditions and unbalanced server loads in edge computing environments. To address the issue of Q-value overestimation in traditional DQN, DDQN is introduced to improve scheduling accuracy in these environments. For example, the work~\cite{zeng2023improved} integrates PSO with DDQN to minimize task scheduling time in edge computing, thereby improving resource utilization efficiency and overall QoS. Additionally, the work~\cite{zhang2018double} leverages DDQN to achieve energy-efficient task scheduling at the network edge by incorporating dynamic voltage and frequency scaling (DVFS) mechanisms, where the evaluation network calculates Q-values for various DVFS configurations, and the target network produces estimated Q-values to guide parameter optimization.

\subsubsection{Policy-Based DRL for Task Scheduling}
Value-based DRL methods have achieved notable progress in task scheduling. However, their dependence on value function estimation often restricts their performance in complex and dynamic scenarios, resulting in challenges such as value overestimation and suboptimal convergence. To address these limitations, policy-based DRL methods offer a promising alternative by directly optimizing scheduling policies. Among these, PPO has shown considerable potential in tackling task scheduling challenges within cloud environments. For example, in the work~\cite{jin2021optimal}, an enhanced PPO algorithm is proposed, integrating priority rules to manage task scheduling with deadlines, particularly in scenarios characterized by random task arrivals and renewable energy integration. This approach aims to minimize both service costs and latency, demonstrating the adaptability of PPO to dynamic and unpredictable scheduling environments. Moreover, the work~\cite{yang2022cloud} utilizes PPO to train agents, which enhances energy efficiency and system performance for cloud task scheduling. In hybrid environment, PPO has proven to be a powerful tool for optimizing task scheduling. The work~\cite{zhao2021deep} introduces an advanced task scheduling framework based on PPO, designed to maximize renewable energy utilization while strictly adhering to deadline requirements in hybrid cloud environments. This framework leverages PPO to enable real-time workload shifting and decision-making, optimizing key metrics such as operational cost, makespan, and resource utilization. In addition to PPO, several studies have explored DDPG-based techniques for task scheduling in cloud computing environment. For instance, in the work~\cite{ran2019slas}, a DDPG-based task scheduling technique is introduced to tackle load balancing and SLA assurance, with a particular focus on optimizing energy consumption in data centers. Expanding on this, the work in~\cite{zhao2022performance} extends DDPG to handle large-scale and heterogeneous cloud workloads, aiming to optimize both response time and cost. This approach leverages a dual reward model to enable agents to effectively learn optimal scheduling policies under complex workload conditions. 

Various policy-based DRL approaches have been proposed to address complexities for task scheduling in edge computing effectively. For example, the work~\cite{zhang2024etpam} leverages the Soft AC algorithm to optimize task scheduling within edge-cloud environments. Additionally, this approach integrates Graph Convolutional Networks (GCN) to capture complex dependencies among tasks through graph-based representations, which are essential for modeling interdependencies in edge-cloud environments. Moreover, the work~\cite{lu2024a2c} employs the A2C algorithm to dynamically schedule tasks in response to fluctuations in edge environments, with the objective of maximizing server resource utilization while minimizing task rejection rates. In another notable approach, the work~\cite{tuli2020dynamic} introduces a task scheduling framework based on the A3C algorithm. This framework incorporates Residual Recurrent Neural Networks (R2N2) to update model parameters in real time, enabling effective adaptation to the stochastic nature of edge-cloud environments. Several works have explored the potential of DDPG in addressing task scheduling challenges in edge computing environments. For example, the work~\cite{zhang2019online} utilizes a DDPG-based algorithm that adjusts network structures to accommodate the discrete action space of edge environments. This approach is specifically designed to enhance scheduling efficiency by maximizing system utility while minimizing cumulative operational costs, effectively meeting the demands of dynamic and resource-constrained edge computing settings. Furthermore, the work~\cite{chen2022dynamic} presents an advanced dynamic task scheduling framework based on the DDPG algorithm, tailored for edge-cloud Internet of Things (IoT) systems. This framework proficiently manages service migration while meeting stringent latency constraints, ensuring optimal performance in highly dynamic settings.  

\subsubsection{Multi-Agent DRL for Task Scheduling}
The application of MADRL in task scheduling has gained significant attention in recent years~\cite{balla2021reliability, tang2022distributed}. MADRL leverages the collaborative capabilities of multiple agents, facilitating dynamic task scheduling while overcoming the scalability and complexity issues that often hinder traditional single-agent approaches~\cite{gergely2024multi}. In cloud environment, for instance, the work~\cite{balla2021reliability} utilizes a MADRL framework for task scheduling, focusing on mitigating total system failure risk and resolving the single point of failure challenge. Additionally, the work~\cite{jung2021orchestrated} presents a cloud-assisted MADRL scheduling framework to optimize task scheduling and energy management in a Unmanned Aerial Vehicle (UAV) charging network. This framework employs a centralized orchestration manager to coordinate energy sharing and scheduling, achieving efficient and adaptive energy distribution across multiple charging stations. Moreover, the work~\cite{gergely2024multi} proposes a general-purpose MADRL framework designed to learn optimal collaborative task scheduling policies. This framework demonstrates the ability to adapt to varying workload demands and effectively manage resource allocation.

Beyond cloud environments, MADRL has been increasingly applied to edge computing, where challenges such as resource heterogeneity, dynamic network conditions, and fluctuating task requirements are prevalent. In these environments, MADRL enables distributed decision-making among edge nodes, improving system efficiency and responsiveness. For instance, the work~\cite{zhang2023multi} introduces a Value-Decomposition Multi-Agent DQN (VD-MADQN) for real-time scheduling of user requests within edge networks. The proposed approach enables edge nodes to independently make scheduling decisions based on localized information while leveraging centralized training to foster cooperative strategies. Additionally, the work~\cite{tang2022distributed} formulates task scheduling in serverless edge computing network as a partially observable stochastic game (POSG). By employing a dueling double deep recurrent Q-network (D3RQN) algorithm, this MADRL framework empowers each edge computing node to autonomously make scheduling decisions, effectively utilizing local observations while aligning with global optimization objectives.

In addition to value-based approaches, recent advancements in MADRL have incorporated policy-based methods to further optimize task scheduling in edge computing. For example, the work~\cite{xu2021multi} proposes a MADRL-based framework built on AC, tailored for distributed transmission within collaborative cloud-edge environments. This approach focuses on joint user scheduling and beam selection, aiming to minimize long-term network delay while maintaining adherence to QoS constraints. In large-scale industrial IoT applications, the work~\cite{zhang2024lsia3cs} introduces a collaborative MADRL framework using the A3C algorithm. This framework allows agents to adapt dynamically to changing conditions, fostering cooperation among agents, and improving both convergence and system stability. Furthermore, the work~\cite{li2024load} develops a competitive multi-agent Attention-Communication AC (MA3C) to support diverse task types within cloud-edge-end collaborative systems. This approach utilizes attention mechanisms for agents to focus on relevant information from other agents in a partially observable environment, thereby enhancing load balancing and overall system efficiency. Similarly, the work~\cite{li2023task} employs a multi-agent PPO (MAPPO) framework for task scheduling in distributed edge computing networks. This framework integrates specific adaptations for edge environments, including state abstraction via heterogeneous graph attention networks (HAN) to capture complex inter-agent semantics and action decomposition for efficient task selection. To address the nonstationarity challenges inherent in heterogeneous edge computing, the work~\cite{niu2023multiagent} develops a multi-agent meta-PPO algorithm. By leveraging meta-learning techniques, this approach accelerates convergence and improves overall system efficiency and stability under dynamic conditions. Lastly, the work~\cite{gong2021multi} provides a comprehensive analysis of two scenarios. In single-edge settings, a DRL-based framework is employed for collaborative task scheduling, while in multi-edge scenarios, the MADDPG algorithm is applied to minimize energy consumption and latency, ensuring efficient resource utilization across edge nodes. 

% Fog computing, serving as a bridge between cloud and edge computing, addresses several challenges of the traditional cloud computing paradigm, such as high latency and security vulnerabilities. The work~\cite{fellir2020multi} proposes a MADRL task scheduling strategy tailored for cloud-fog computing platforms, emphasizing the prioritization of critical tasks. The proposed method demonstrates significant adaptability in fog computing environments, facilitating dynamic and flexible resource allocation based on task priority.

% These studies highlight the strengths of multi-agent reinforcement learning and deep reinforcement learning in addressing the specific challenges of fog and edge computing environments, enhancing resource management, and optimizing task execution in distributed, large-scale systems.

\subsubsection{Advanced DRL for Task Scheduling}
To overcome the inherent complexities for task scheduling in cloud computing environments, recent research has incorporated advanced techniques into DRL frameworks, pushing the boundaries of task scheduling optimization. For instance, the work~\cite{huang2021deep} proposes a hybrid framework that integrates adversarial imitation learning with the Dueling Double Deep Q-Network (D3QN) algorithm for cloud task scheduling. By utilizing adversarial imitation learning to store high-reward job trajectories as expert demonstrations, this approach directly guides the DRL agent, enhancing both policy optimization and scheduling performance. The work~\cite{rjoub2019deep} presents a DRL framework integrated with LSTM networks to optimize VM scheduling in big data analytics. The LSTM network captures long-term dependencies between tasks and resource demands, enhancing scheduling efficiency and reducing execution costs. The work~\cite{li2023batch} proposes a distributional RL-based approach for load balancing in cloud computing environments, with a focus on batch task scheduling. By employing quantile regression, this framework effectively distributes computational loads across resources, adapting dynamically to fluctuations in job requirements and cluster states. Additionally, the work~\cite{mao2024dl} introduces a Double-Level DRL approach that includes an Interactive Training Strategy (ITS), designed to boost adaptability and scalability.

As edge computing environments expand in scale and complexity, advanced DRL techniques are increasingly applied to address the demands of dynamic, high-dimensional scheduling tasks. For instance, the work~\cite{tang2020representation} proposes a task scheduling framework that combines DDQN with representation learning to handle the complexities of dynamic edge environments. By reducing the dimensionality of nodes and tasks, the representation learning component enables efficient high-dimensional data processing, thereby enhancing the speed and accuracy of DRL-based decision-making in edge computing environments. Additionally, the work~\cite{zeng2023improved} improves DDQN efficiency by incorporating a pre-training phase with PSO, which provides a near-optimal initialization for task scheduling. This hybrid approach accelerates the initial learning phase of DDQN and addresses overestimation issues in DQNs by decoupling action selection from target Q-value computation. In a parallel direction, the work~\cite{qi2020scalable} introduces a scalable multi-task DRL framework for parallel task scheduling, which leverages multi-task learning to efficiently manage high-dimensional action spaces and concurrently optimize multiple tasks. This approach reduces computational overhead while enhancing resource utilization in edge environments.

% Subsequently, the work~\cite{zhang2017energy} introduces an energy-efficient scheduling method for real-time systems. Focusing on periodic tasks in embedded real-time systems, this DQN-based approach combines stacked autoencoders with Q-learning to dynamically adjust system voltage and frequency. The model learns optimal scheduling strategies for different system states, enabling the selection of the most energy-efficient task scheduling while ensuring timely task completion, thereby minimizing processor energy consumption.
\begin{table*}[!t]
\centering
\caption{Summary of typical works on DRL-based workflow scheduling}
\label{tab:workflow_scheduling}
    \setlength{\tabcolsep}{3.2mm}{
    \resizebox{\linewidth}{!}{
    \large 
    \begin{tabular}{cccccccc}
        \toprule
        % \multicolumn{1}{c}{Reference} & \multicolumn{1}{c}{Method} & \multicolumn{1}{c}{Others} & \multicolumn{1}{c}{Environment} & \multicolumn{4}{c}{Optimization objective} \\ 
        % & & & & Makespan & Cost & Energy consumption & Others \\ 
        \multicolumn{1}{c}{\multirow{2}{*}{Reference}} & \multicolumn{1}{c}{\multirow{2}{*}{Method}} & \multicolumn{1}{c}{\multirow{2}{*}{Others}} & \multicolumn{1}{c}{\multirow{2}{*}{Environment}} & \multicolumn{4}{c}{Optimization objective} \\
        & & & & Makespan & Cost & Energy consumption & Others \\
        \midrule      
        ~\cite{tong2020scheduling} & Deep Q-learning & - & Cloud computing & \checkmark & - & - & Load balance \\
        ~\cite{pan2024deep} & DQN & - & Cloud computing & \checkmark & - & - & Resource utilization \\ 
        ~\cite{kintsakis2019reinforcement} & DQN & SA & Cloud computing & \checkmark & \checkmark & - & Task failures; communication costs \\
        ~\cite{gu2024cost} & DQN & SA & Cloud computing & \checkmark & \checkmark & - & - \\
        ~\cite{zhang2023cost} & DQN & GA & Cloud computing & \checkmark & \checkmark & - & - \\
        ~\cite{dong2023deep} & DDQN & - & Cloud computing & \checkmark & - & - & Resource usage \\
        ~\cite{li2022weighted} & DDQN & - & Cloud computing & \checkmark & \checkmark & - & - \\
        ~\cite{chen2023collaborative} & D3QN & - & Cloud computing & \checkmark & \checkmark & - & Fairness and continuity \\
        ~\cite{zhang2023data} & DQN & - & Cloud computing & \checkmark & - & - & Load balance \\
        ~\cite{han2022edgetuner} & DQN & - & Edge-Cloud collaboration & \checkmark & - & - & - \\
        ~\cite{zheng2022deep} & DQN & - & Edge-Cloud collaboration & \checkmark & - & - & VM utilization; failed tasks rate \\
        ~\cite{qian2020workflow} & DQN & - & Edge-Cloud collaboration & \checkmark & - & - & Network throughput \\
        ~\cite{cai2024task} & DDQN & LSTM & Edge-Cloud collaboration & \checkmark & - & \checkmark & Utility \\
        ~\cite{xie2022workflow} & D3QN & - & Edge computing & \checkmark & \checkmark & \checkmark & - \\
        ~\cite{dong2021workflow} & AC & - & Cloud computing & \checkmark & - & - & - \\  
        ~\cite{dong2021deep} & AC & LSTM & Cloud computing & \checkmark & - & - & - \\
        ~\cite{koslovski2024dag} & AC & - & Cloud computing & - & - & - & Bounded slowdown; resource utilization \\
        ~\cite{wang2023reinforcement} & A2C & - & Cloud computing & - & - & \checkmark & Carbon emission \\
        ~\cite{xue2023towards} & PPO & GCN & Cloud computing & \checkmark & - & - & Job latency \\
        ~\cite{peng2022lore} & PPO & GCN & Cloud computing & \checkmark & - & - & Resource utilization \\
        ~\cite{lin2024spotdag} & PPO & Self-attention mechanism & Cloud computing & - & \checkmark & - & - \\
        ~\cite{yang2023effective} & DDPG & - & Cloud computing & \checkmark & - & - & - \\
        ~\cite{jayanetti2022deep} & AC & - & Edge-Cloud collaboration & \checkmark & - & \checkmark & - \\
        ~\cite{li2023componentized} & AC & GCN & Edge-Cloud collaboration & \checkmark & - & - & Energy cost; network traffic; load balance \\  
        ~\cite{mounesan2024reinforcement} & A3C & - & Edge-Cloud collaboration & - & - & - & Task rejection rate; resource utilization \\
        ~\cite{zhu2024learning} & PPO & GNN & Edge-Cloud collaboration & \checkmark & - & - & Migration time; energy cost \\   
        ~\cite{wang2024deep} & PPO & - & Edge-Cloud collaboration & \checkmark & \checkmark & - & Load balance \\
        ~\cite{zhang2023time} & PPO & - & Edge computing & \checkmark & - & - & Resource utilization \\
        ~\cite{zhu2022workflow} & DDPG & - & Edge computing & \checkmark & - & - & - \\
        ~\cite{asghari2020cloud} & MADRL & - & Cloud computing & - & \checkmark & \checkmark & Load balance; resource utilization \\ 
        ~\cite{limulti} & MADRL & - & Cloud computing & \checkmark & \checkmark & - & - \\   
        ~\cite{jayanetti2024multi} & MADDPG & - & Cloud computing & \checkmark & - & \checkmark & - \\     
        ~\cite{jayanetti2024deep} & MADRL & - & Cloud computing & \checkmark & \checkmark & - & Execution interruptions \\     
        ~\cite{duan2024telemetry} & MADRL & - & Edge computing & \checkmark & - & \checkmark & Success rate; resource utilization \\
        ~\cite{huang2023digital} & MADRL & - & Edge computing & \checkmark & - & \checkmark & Success rate; load balance \\
        ~\cite{zhao2024multi} & MADRL & - & Edge computing & \checkmark & - & - & - \\
        % ~\cite{zhang2023hybrid} & MADRL & - & Edge computing & \checkmark & \checkmark & \checkmark & system reliability \\
        ~\cite{wang2022adaptive} & Depth-First-Search Coalition RL & - & Cloud computing & \checkmark & - & - & - \\
        ~\cite{ding2024transformer} & DQN & Transformer & Cloud computing & \checkmark & \checkmark & \checkmark & - \\     
        ~\cite{liu2024ga} & DDQN & GNN & Cloud computing & \checkmark & - & - & - \\
        ~\cite{long2022fault} & DQN & Primary backup & Edge computing & \checkmark & - & - & - \\
        ~\cite{mahapatra2024quantum} & DQN & QML & Edge-Cloud collaboration & \checkmark & \checkmark & - & Power utilization; violation of delay  \\
        ~\cite{wang2024tf} & DQN & Transformer & Edge-Cloud collaboration & \checkmark & \checkmark & \checkmark & Weighted cost \\
       
        % ~\cite{rjoub2019deep} & DQN & LSTM & cloud computing & - & - & - & CPU usage cost; RAM memory usage cost \\      
        % ~\cite{zhang2017energy} & PPO & & edge cloud & 1 & 1 & 1 \\
        \bottomrule
        \end{tabular}}}
\end{table*}

\subsection{DRL-Based Workflow Scheduling}

Workflow applications are increasingly prevalent in cloud environments due to their effectiveness in addressing complex challenges in scientific research, such as astronomy, bioinformatics, and seismology~\cite{deelman2015pegasus}. A workflow application typically consists of multiple computational tasks organized as a DAG, where nodes represent individual tasks and edges indicate dependencies among these tasks. The inherent complexity of such a structure makes workflow scheduling an NP-hard combinatorial optimization problem. Recent research has reformulated workflow scheduling as a discrete-time control problem, leveraging DRL techniques to develop more adaptive and scalable scheduling frameworks~\cite{tong2020scheduling, pan2024deep, kintsakis2019reinforcement}. Below, we provide a detailed analysis of these studies and summarize them in Table~\ref{tab:workflow_scheduling}.

\subsubsection{Valued-Based DRL for Workflow Scheduling} 

Valued-based DRL methods, represented by DQN, have demonstrated their effectiveness in addressing workflow scheduling challenges within cloud computing environments. For example, the work~\cite{tong2020scheduling} employs DQN to handle DAG-based tasks in a cloud computing environment, with a primary focus on reducing makespan variance and improving load balance. Furthermore, DQN has been successfully applied in real-time workflow scheduling. In a dynamic environment with multiple real-time workflows, the work~\cite{pan2024deep} applies DQN to assign each task in workflows to a suitable VM, thereby minimizing workflow makespan and maximizing resource utilization. Similarly, the work~\cite{kintsakis2019reinforcement} leverages DQN to optimize real-time workflow scheduling in cloud environments, aiming to reduce execution time and dynamically manage resource allocation. To further enhance the performance of DRL in workflow scheduling, heuristic algorithms have been integrated with DQN, resulting in hybrid models that accelerate convergence and mitigate the risk of local optima. For instance, the work~\cite{gu2024cost} introduces a hybrid model combining DQN with SA for cost-sensitive workflow scheduling in cloud environments. In this approach, SA optimizes the task execution sequence, while DQN learns optimal scheduling policies in dynamic environments, ultimately improving resource utilization and reducing scheduling costs. Likewise, the work~\cite{zhang2023cost} integrates GA with DQN, where GA is responsible for optimizing VM execution plans through a global search strategy, effectively reducing the exploration space of DQN. This integration minimizes execution costs and response times, thereby enhancing scheduling efficiency.

Building on the success of DQN in cloud-based workflow scheduling, its variants, such as DDQN~\cite{dong2023deep} and D3QN~\cite{chen2023collaborative} have significantly expanded its decision-making capabilities, enabling more effective solutions to complex scheduling challenges. For instance, the work~\cite{dong2023deep} proposes a DDQN-based scheduling framework that mitigates Q-value overestimation by decoupling action selection from evaluation, thus minimizing task completion time and resource consumption. Similarly, the work~\cite{li2022weighted} introduces a weighted DDQN approach that incorporates adaptive dynamic coefficients to balance Q-value overestimation in DQN with the underestimation in DDQN. This approach enables simultaneous optimization of execution time and cost. Moreover, the work~\cite{chen2023collaborative} introduces a D3QN-based collaborative scheduling strategy for heterogeneous workflows in cloud. By incorporating dueling and double Q-learning mechanisms, this strategy effectively addresses Q-value overestimation while optimizing workflow makespan, cost, fairness, and continuity. Additionally, the work~\cite{zhang2023data} incorporates a bias correction mechanism to further alleviate Q-value overestimation, resulting in optimized task execution times and balanced load distribution across multi-cloud environments.

Beyond cloud computing, recent advancement in valued-based DRL methods have driven the development of more sophisticated workflow scheduling frameworks tailored to the unique requirements of edge computing environments. For example, the work~\cite{han2022edgetuner} presents a DQN-based workflow scheduling framework for dynamic edge-cloud workloads, utilizing Implicit Quantile Networks (IQN) to enhance robustness and Ape-X for distributed experience replay. Trained offline using a cluster simulator, this framework adaptively selects the most suitable scheduling strategies to improve resource allocation and minimize response times under fluctuating conditions. Moreover, the work~\cite{zheng2022deep} develops a DQN-based scheduling algorithm that aims to balance workload distribution, reduce service time, and minimize task failure rates by optimizing task allocation across edge and cloud resources. Within IoT applications, the work~\cite{qian2020workflow} utilizes DQN to develop a workflow-aided IoT paradigm that integrates intelligence edge computing. This method automates the handling of task dependencies, improving VM allocation and minimize network delay. To address the limitations of DQN, some studies have explored advanced DQN variants to enhance scheduling efficiency and robustness under dynamic edge-cloud conditions. For example, the work ~\cite{cai2024task} presents a hierarchical workflow scheduling framework tailored for collaborative cloud-edge-end computing environments with a DDQN approach. Furthermore, the work~\cite{xie2022workflow} employs a D3QN-based method that addresses Q-value overestimation by leveraging two Q-networks updated in an alternating manner. This method effectively optimizes key performance metrics such as execution time, energy consumption, and operational costs within edge computing environments.
% This framework utilizes a Large-Small Resource Tree (LST) model at the edge layer for efficient resource management and leverages an LSTM-based multi-granularity task decomposition algorithm to predict resource availability, thereby optimizing task granularity across the multi-tiered network.

\subsubsection{Policy-Based DRL for Workflow Scheduling}
% Policy-based DRL offers significant advantages over value-based methods for workflow scheduling by directly optimizing scheduling policies. Unlike value-based DRL, which relies on value function estimation, policy-based DRL efficiently manages intricate task dependencies and dynamic resource variations, particularly in high-dimensional task environments, allowing for faster adaptation to the dynamic arrival of tasks and enhancing workflow execution efficiency and resource utilization through timely policy updates. 

Policy-based DRL methods offer significant advantages for workflow scheduling by directly optimizing scheduling policies. These methods effectively manage dynamic resource variations and adapt to changes in task arrivals, improving workflow execution efficiency and resource utilization through timely policy updates. The AC framework in policy-based DRL has been effectively applied to tackle the challenges of workflow scheduling in cloud computing. For example, the work~\cite{dong2021workflow} presents an AC framework to manage the complexities of cloud workflow scheduling. This method integrates a P-Network model for task prioritization prediction, supplemented by a heuristic algorithm that facilitates server selection based on task precedence and computational complexity considerations. Similarly, the work~\cite{dong2021deep} proposes a dynamic workflows scheduling approach based on the AC model, with the objective of minimizing makespan. This approach utilizes an extended pointer network and the LSTM-based encoder to handle the dependency structure of workflows, enabling adaptive task allocation that improves efficiency under dynamic resource conditions. Additionally, the work~\cite{koslovski2024dag} introduces the scheduling of DAG-based workflows using the AC method. This approach considers both the dependency structure within DAGs and the resource capacities of cloud data centers, aiming to improve scheduling decisions and optimize resource usage. Moreover, the work~\cite{wang2023reinforcement} proposes an eco-fridenly A2C-based framework for workflow scheduling in federated cloud environment. The objective is to minimize energy consumption and carbon emissions by intelligently selecting data centers for workflow task execution. 

The PPO algorithm has also been widely adopted to enhance workflow scheduling efficiency, advancing beyond the standard AC framework by introducing a stable policy optimization mechanism that constrains update magnitudes in cloud environment. For example, the work~\cite{xue2023towards} proposes a PPO-based workflow scheduler for cluster, aiming to minimize job latency and makespan while prioritizing critical tasks. By leveraging the GCN to extract workflow features, the proposed scheduler optimizes resource allocation across multiple queues. Similarly, by incorporating Monte Carlo Tree Search (MCTS) and GCN into PPO, the work~\cite{peng2022lore} focuses on reducing workflow makespan while increasing resource utilization in the cloud environments. In heterogeneous cloud environments, the work~\cite{lin2024spotdag} utilizes PPO for scheduling workflow, with the objective of minimizing instance costs and ensuring timely completion of all tasks. This approach incorporates a self-attention mechanism for effective feature extraction and employs a mask layer to prevent illegal actions, thereby enhancing the stability and convergence speed of the learning process. In addition to PPO, DDPG has also been explored for workflow scheduling. For example, the work~\cite{yang2023effective} proposes a workflow scheduling policy based on the DDPG algorithm to minimize computation latency in distributed cloud computing systems.

The heterogeneity and dynamic characteristics of edge-cloud environments present additional challenges for workflow scheduling, further driving the adoption of policy-based DRL methods.  For example, the work~\cite{jayanetti2022deep} proposes a hybrid AC model for optimizing workflow scheduling in edge-cloud environments, aiming to minimize energy consumption while ensuring timely task execution. This approach integrates multiple actor networks with a single critic network, effectively supporting hierarchical action spaces that distinguish between edge and cloud nodes. Expanding on similar principles, the work~\cite{li2023componentized} proposes a framework based on the AC model to optimize the scheduling of componentized tasks within cloud-edge environments. This framework integrates a Graph Neural Network (GNN)-enhanced DRL, utilizing a combination of GCN and Directed GCN to effectively capture and represent task and network graph information, optimizing multiple objectives such as system latency, energy cost, network traffic, and load balancing. Additionally, the work~\cite{mounesan2024reinforcement} utilizes the A3C algorithm for data-intensive workflow scheduling in a volunteer edge-cloud environment, considering workflow QoS requirements, security specifications, and the resource preferences of volunteer nodes (VNs). The A3C-based model employs both actor and critic networks within a parallel learning architecture involving multiple workler agents, enabling efficient adaptation to dynamic and heterogeneous environments.

PPO has also been effectively utilized to address workflow scheduling challenges in edge-cloud environments, offering robust solutions for managing dynamic and time-sensitive tasks. For example, the work~\cite{zhu2024learning} explores a PPO-based workflow scheduling in the edge-cloud computing environment with continuous task arrivals, leveraging a GNN-based workflow embedding to capture latent task dependency information. The framework introduces an intrinsic reward mechanism to provide immediate feedback and improve scheduling decisions dynamically, ultimately aiming to minimize makespan and enhance energy utilization. In addition, the work~\cite{wang2024deep} proposes a PPO-based IoT application scheduling algorithm, which aims to optimize system load balancing and response time in edge and fog computing environments. To further enhance scheduling in time-sensitive and resource-constrained edge environments, the work~\cite{zhang2023time} develops a workflow scheduling framework based on PPO, incorporating a self-critic mechanism for improved decision-making efficiency and transformer-based neural networks for better sub-task feature processing. Furthermore, the DDPG algorithm has been employed by the work~\cite{zhu2022workflow} for workflow scheduling in the edge network, focusing on minimizing makespan. This framework integrates critical path analysis-based dynamic task sorting and a path quality indicator for multi-path routing, effectively coordinating computing resources and managing interdependent tasks.

\begin{figure*}[!t]
\centering
\includegraphics[width=1\linewidth]{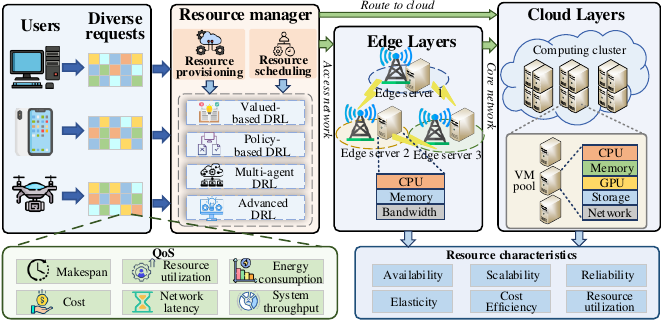}
\caption{The system architecture of resource management in cloud computing}
\label{fig:resource_management}
\end{figure*}

\subsubsection{Multi-Agent DRL for Workflow Scheduling}
MADRL has demonstrated significant potential for tackling workflow scheduling challenges. In cloud environments, MADRL has been widely adopted for workflow scheduling, with the goal of solving complex scheduling challenges and improving overall system performance. For example, the work~\cite{asghari2020cloud} proposes a workflow scheduling framework that leverages MADRL to manage multiple online scientific workflows. This framework integrates cooperative Q-learning with Markov game theory to optimize both resource provisioning and task scheduling, effectively balancing the objectives of reducing energy consumption, minimizing user costs, and distributing workloads evenly across resources for both users and cloud providers. Additionally, the work~\cite{limulti} proposes a MADRL framework based on DQN for multi-objective workflow scheduling in cloud environment. This framework focuses on optimizing workflow makespan and cost, dynamically adapting to different workflow types and VM configurations to achieve effective scheduling. Moreover, the work~\cite{jayanetti2024multi} presents a MADRL approach based on AC, specifically designed to optimize renewable energy usage in workflow scheduling across distributed cloud data centers. By employing a hierarchical approach, the framework features a global RL agent to assign tasks to data centers, while local RL agents assign tasks to individual nodes. This setup effectively manages the complexities of partial observability and distributed environments. Additionally, the work~\cite{jayanetti2024deep} utilizes a MADRL framework based on PPO to minimize workflow execution costs by optimizing the use of both preemptible and on-demand instances. This approach features multiple actor networks coordinated under the guidance of a centralized critic network, leveraging a hierarchical action space to ensure an optimal balance between resource utilization and operational efficiency in cloud environments.

In edge-cloud environments, the use of MADRL has demonstrated considerable potential in optimizing workflow scheduling under dynamic conditions. For instance, the work~\cite{duan2024telemetry} presents a telemetry-aided cooperative MADRL framework based on DQN for workflow scheduling in edge computing environment. By utilizing telemetry data for real-time decision-making, this framework aims to improve resource allocation efficiency and workflow scheduling across edge servers and programmable switches, ultimately enhancing the performance of workflow execution under dynamic network conditions. Moreover, the work~\cite{huang2023digital} integrates Digital Twin (DT) technology with Evolutionary Selection Multi-Agent Reinforcement Learning (ES-MARL) to address workflow scheduling challenges in large-scale MEC environments. In this approach, DT assists in maintaining up-to-date network states for centralized training, while distributed agents make local scheduling decisions. The use of the Multi-Agent Transformer (MAT) algorithm, combined with evolutionary selection, improves training efficiency and fosters effective collaboration among agents. In addition, the work~\cite{zhao2024multi} proposes a workflow scheduling approach based on MADRL for vehicular edge computing (VEC) environments. This method models the task offloading problem as a potential game and leverages DRL to enhance collaboration between vehicles and Roadside Units (RSUs), resulting in improved offloading decisions and resource utilization.

\subsubsection{Advanced DRL for Workflow Scheduling}
Recent advances in DRL have driven significant progress in workflow scheduling by enhancing workflow representations~\cite{liu2024ga}, integrating hybrid methods like quantum machine learning (QML)~\cite{mahapatra2024quantum}, and improving scheduling efficiency through sophisticated neural architectures such as Transformers~\cite{ding2024transformer, wang2024tf}, etc. These advancements have transformed both cloud and edge computing environments. For cloud environment, advanced DRL approaches have introduced effective solutions for handling complex scheduling demands. For example, the work~\cite{wang2022adaptive} presents an adaptive multi-workflow scheduling framework for cloud computing environments, employing a Depth-First-Search Coalition Reinforcement Learning (DFSCRL) policy. This policy integrates physical machines (PMs) coalition formation with Q-learning to determine the optimal bundle of VM instances, leading to improved resource efficiency, reduced costs, and enhanced workflow reliability. Moreover, the work~\cite{ding2024transformer} extends the DRL by proposing a transformed-enhanced DQN approach for efficient large-scale dynamic workflow scheduling in heterogeneous cloud environments. By integrating transformer models, this approach effectively handles complex dependencies, uncertainties in task execution times, and dynamic resource availability, thereby optimizing system performance across diverse workflow scenarios. Additionally, the work~\cite{liu2024ga} proposes a GNN-augmented DDQN framework for workflow scheduling in dynamic cloud environments. This framework employs a multi-head Graph Attention Network (GAT) to extract features of subtasks, accounting for both predecessor and successor relationships. 

For edge computing, advanced DRL methods have addressed the unique challenges posed by distributed and resource-constrained environments. For example, the work~\cite{long2022fault} presents a fault-tolerant workflow scheduling method leveraging the Primary-Backup (PB) strategy in combination with DQN. This method enhances workflow reliability by mitigating resource and link failures, which are common in distributed edge environments. Moreover, the work~\cite{mahapatra2024quantum} introduces a novel hybrid approach that integrates DQN with QML for workflow scheduling across edge, fog, and cloud layers. By utilizing QML principles such as superposition and entanglement, the proposed method effectively addresses challenges like power consumption, delay minimization, and service cost reduction in collaborative edge-cloud environments. Additionally, the work~\cite{wang2024tf} introduces a distributed DRL framework enhanced by transformer for workflow scheduling in edge-cloud environments. By incorporating Prioritized Experience Replay (PER) and transformer layers into the DRL architecture, the framework effectively reduces high exploration costs and captures long-term dependencies among tasks, leading to better system performance.

\begin{table*}[!t]
\centering
\caption{Summary of typical works on DRL-based resource provisioning}
\label{tab:literature_resource_provisioning}
    \setlength{\tabcolsep}{3.2mm}{
    \resizebox{\linewidth}{!}{
    \begin{tabular}{cccccccc}
        \toprule
        % \multicolumn{1}{c}{Reference} & \multicolumn{1}{c}{Method} & \multicolumn{1}{c}{Others} & \multicolumn{1}{c}{Environment} & \multicolumn{4}{c}{Optimization objective} \\
        % & & & & Makespan & System Cost & Energy consumption & Others \\
        % \midrule
         \multicolumn{1}{c}{\multirow{2}{*}{Reference}} & \multicolumn{1}{c}{\multirow{2}{*}{Method}} & \multicolumn{1}{c}{\multirow{2}{*}{Others}} & \multicolumn{1}{c}{\multirow{2}{*}{Environment}} & \multicolumn{4}{c}{Optimization objective} \\
        & & & & Makespan & Cost & Energy consumption & Others \\
        \midrule
        ~\cite{qingADRLBasedResourceAutonomicProvisioningApproach2021} & DQN & - & Cloud computing & - & \checkmark & - & SLA requirements \\
        ~\cite{tuliCILPCoSimulationBasedImitation2023} & DQN & - & Cloud computing & \checkmark & \checkmark & \checkmark & SLA requirements \\
        ~\cite{zhu2022deep} & DDQN & - & Cloud computing & - & \checkmark & - & Spectrum consumption \\
        % ~\cite{panwarRLPRAFReinforcementLearningBased2024} & Q-learning & - & cloud computing & - & \checkmark & - & SLA requirements \\
        ~\cite{sami2021ai} & DQN & - & Edge computing & - & \checkmark & - & Resource utilization \\
        ~\cite{zhu2022drl} & DQN & - & Edge-Cloud collaboration & - & - & - & Initial delay; blocking probability \\
        ~\cite{faraji-mehmandarSelflearningApproachProactive2022} & DQN & - & Edge computing & \checkmark & \checkmark & \checkmark & Resource utilization \\
        ~\cite{santosResourceProvisioningFog2021} & DQN & - & Edge computing & - & - & \checkmark & - \\
        ~\cite{chen2021adaptive} & A3C & - & Cloud computing & - & - & \checkmark & QoS \\
        ~\cite{funikaAutomatedCloudResources2023} & PPO & - & Cloud computing & - & \checkmark & - & Resource utilization \\
        ~\cite{zhang2021you} & DDPG & - & Cloud computing & - & - & - & Blocking probability \\
        ~\cite{baghbanEdgeAIIoTRequest2024} & Actor-Critic & - & Edge computing & - & \checkmark & - & Execution delay \\
        ~\cite{guo2019trusted} & A3C & - & Edge-Cloud collaboration & \checkmark & \checkmark & - & QoS \\
        ~\cite{sIntelligentResourceProvisioning2023} & PPO & - & Edge computing & - & \checkmark & \checkmark & Resource utilization \\
        ~\cite{chenDeepReinforcementLearning2021} & DDPG & - & Edge computing & - & \checkmark & - & - \\
        ~\cite{zhang2021dynamic} & DDPG & - & Edge computing & - & - & - & Resource utilization \\
        ~\cite{jyotiDynamicProvisioningResources2020} & MADRL & - & Cloud computing & \checkmark & \checkmark & \checkmark & Resource utilization; load balancing \\
        ~\cite{oudaa2021agent} & MADRL & QR-DQN & Cloud computing & - & - & \checkmark & - \\
        ~\cite{asghariCombinedUseCoral2022} & MADRL & CRO & Cloud computing & - & - & \checkmark & - \\
        ~\cite{zhangDistributedMultiCloudMultiAccess2021} & MARL & - & Edge-Cloud collaboration & - & - & - & System latency \\
        ~\cite{li2023adaptive} & MADDPG & - & Edge computing & - & - & - & Rraining time cost \\
        ~\cite{asimdynamic} & MAA2C & - & Edge computing & - & \checkmark & - & Request delay \\
        ~\cite{asghariTaskSchedulingResource2021} & SARSA & GA & Cloud computing & \checkmark & - & - & Resource utilization; load balancing \\
        ~\cite{xuGenerativeAIenabledQuantum2024} & MADRL & Generative AI & Cloud computing & - & \checkmark & \checkmark & Execution latency \\
        ~\cite{shahidinejadJointComputationOffloading2020} & Q-learning & LSTM and LA & Edge computing & \checkmark & - & \checkmark & Offloading \\
        ~\cite{heIntelligentProvisioningVirtualized2022} & - & POKTR & Edge computing & - & - & - & TUR \\
        ~\cite{dieyeDRLBasedGreenResource2023} & DQN & MCTS & Edge computing & - & \checkmark & - & QoS \\
        ~\cite{chen2019iraf} & RL & MCTS & Edge computing & - & - & \checkmark & Service latency \\

        \bottomrule
        \end{tabular}}}
\end{table*}

\section{DRL-based Cloud Resource Management}\label{resource_management}
In this section, we review existing works on DRL-based resource provisioning and scheduling for cloud resource management. Resource provisioning focuses on allocating virtualized resources to meet user demands or workload requirements, while resource scheduling involves efficiently assigning these resources to tasks. DRL-based resource management has proven to be an adaptive and efficient solution for addressing the complexities of both provisioning and scheduling. By leveraging DRL techniques, resource managers dynamically optimize utilization, minimize operational overhead, and ensure compliance with QoS requirements across cloud and edge environments. Fig.~\ref{fig:resource_management} illustrates the architecture of a DRL-based resource management system, showcasing its flexibility in adapting to diverse and dynamic user demands.

\subsection{DRL-Based Resource Provisioning}

Resource provisioning is a critical aspect of resource management within cloud computing, with an emphasis on efficiently allocating resources to meet fluctuating demands and optimize performance metrics. In the following, we provide an in-depth analysis of DRL-based resource provisioning strategies, focusing on their deployment within cloud and edge computing environments, while examining their capacity to improve system efficiency, reduce costs, and enhance QoS. Table~\ref{tab:literature_resource_provisioning} offers a summary of these approaches.

\subsubsection{Value-Based DRL for Resource Provisioning}

Value-based DRL approaches are well-suited for optimizing resource provisioning by estimating action values to guide decision-making. DQN has been widely applied to address dynamic resource provisioning challenges, particularly in cloud and containerized environments. For example, the work~\cite{qingADRLBasedResourceAutonomicProvisioningApproach2021} introduces an elastic resource provisioning method that integrates DQN to achieve efficient horizontal scaling for cloud services. By dynamically adjusting the number of VMs in response to fluctuating user demands, this approach minimizes resource wastage and reduces SLA violations, thus improving resource utilization and reducing overall costs. Similarly, the work~\cite{tuliCILPCoSimulationBasedImitation2023} applies DQN to optimize VM configurations based on workload characteristics, aiming to minimize energy consumption, operational costs, and task response time, while simultaneously improving QoS and resource utilization. However, the inherent overestimation bias in DQN can limit its efficacy in resource provisioning. To address this, the work~\cite{zhu2022deep} proposes a DDQN-based solution for resource provisioning in inter-datacenter elastic optical networks. This framework optimizes the deployment and reuse of Virtual Network Functions (VNFs), enabling more efficient management of IT and spectrum resources to meet the demands of VNF service chaining. By decomposing complex VNF service chains (VNF-SCs) into manageable segments and applying an encoding scheme for standardized input lengths, DDQN facilitates efficient processing of diverse provisioning tasks.

In edge computing environments, where resource demands are often dynamic and heterogeneous, valued-based DRL methods have also been widely adopted to address unique provisioning challenges. For example, the work~\cite{sami2021ai} introduces a DQN-driven framework designed to tackle resource provisioning challenges in MEC for 6G networks, enabling efficient, adaptive resource scaling and optimal service placement for IoE services. Similarly, the work~\cite{zhu2022drl} proposes a DQN-based solution for resource provisioning in cloud-edge environments connected via elastic optical networks (EONs). By optimizing both time and spectrum resources, this framework reduces network fragmentation and blocking probability, dynamically adapting to deadline-sensitive demands and improving resource utilization. Furthermore, the work~\cite{faraji-mehmandarSelflearningApproachProactive2022} presents a self-learning DQN-based method for proactive resource and service provisioning in edge computing, targeting reductions in response time, cost, and energy consumption while improving overall resource utilization. To address the overestimation bias in conventional DQN, the work~\cite{santosResourceProvisioningFog2021} introduces a DRL method called D3QN-PER, which combines D3QN with Prioritized Experience Replay for Service Function Chaining Allocation (SFCA) in fog computing. This enhanced method adapts to dynamic network conditions while improving energy efficiency.

\subsubsection{Policy-Based DRL for Resource Provisioning}

Policy-based DRL techniques have emerged as powerful tools for determining optimal resource provisioning decisions, enabling more efficient and adaptive management of resources in dynamic environments. In the context of cloud computing, several studies have explored the application of policy-based DRL methods to enhance resource allocation and management. For instance, the work~\cite{chen2021adaptive} proposes an adaptive resource provisioning framework based on the A3C algorithm. This framework utilizes the policy-based DRL to handle dynamic system states and heterogeneous user demands, improving QoS and energy efficiency in cloud datacenters. Moreover, the work~\cite{funikaAutomatedCloudResources2023} proposes an automated cloud resource provisioning method based on PPO to address the issue of heterogeneous resource management. Specifically, this framework dynamically adjusts the number and types of VMs in a cloud environment, with the goal of enhancing system cost-efficiency and resource utilization by automatically adjusting resource provisioning. Furthermore, the work~\cite{zhang2021you} introduced a DDPG-based service framework for resource provisioning in datacenter interconnections (DCIs), concentrating on virtual network slicing. This framework utilizes DDPG to assist in pricing and advertising substrate resources across non-overlapping subgraphs of a DCI, allowing tenants to compute their own virtual network embedding (VNE) schemes independently, which reduces resource conflicts and computation time. This tenant-driven approach optimizes resource utilization and cost-effectiveness while improving scalability and efficiency compared to traditional centralized methods.

Similarly, in edge and fog computing environments, policy-based DRL strategies have been increasingly employed to address specific challenges related to resource provisioning and service optimization. For instance, the work~\cite{baghbanEdgeAIIoTRequest2024} introduces the DRL-Dispatcher, which uses an AC algorithm to optimize task scheduling and resource provisioning for IoT requests, focusing on minimizing execution delays and operational costs. To address the complexities of resource provisioning and service migration in a heterogeneous cloud-edge environment supporting IoT applications, this work~\cite{guo2019trusted} utilizes A3C algorithm for trusted and dynamic Service Function Chain (SFC) orchestration. The approach aims to minimize orchestration costs and improve QoS, optimizing resource provisioning in both edge and cloud layers to meet the demands of high-mobility IoT networks. Moreover, PPO has been widely applied to address resource provisioning challenges in edge computing environments. The work~\cite{sIntelligentResourceProvisioning2023} addresses fog computing challenges by utilizing the PPO algorithm to overcome issues such as uneven resource provisioning, suboptimal QoS, and low network bandwidth utilization. This approach aims to reducing latency, minimize deployment costs, and improve resource utilization. Furthermore, DDPG has been employed to tackle similar challenges in resource provisioning. In mobile edge computing, the work~\cite{chenDeepReinforcementLearning2021} proposes a DDPG-based method for computation offloading, which predicts optimal resource provisioning actions to reduce overall system costs. Additionally, the work~\cite{zhang2021dynamic} explores dynamic edge server reservation for connected vehicles in edge computing. This framework employs a DDPG algorithm enhanced with ConvLSTM and an action amender to capture spatio-temporal correlations in resource demand. By adapting server reservation decisions to real-time workload observations, this approach effectively addresses fluctuating resource requirements and minimizes provisioning inefficiencies.

\subsubsection{Multi-Agent DRL for Resource Provisioning}

MADRL has emerged as a powerful approach for optimizing resource provisioning, particularly in distributed systems where coordination among multiple agents is essential. Recent advancements in cloud computing have demonstrated the effectiveness of MADRL in addressing resource allocation and energy efficiency, particularly in tackling the challenges of dynamic provisioning and large-scale optimization. For example, the work~\cite{jyotiDynamicProvisioningResources2020} proposes a dynamic resource provisioning framework in cloud computing, focused on improving load balancing and service brokering through a MADRL approach. Specifically, the MADRL model anticipates user demand to prioritize resource allocation to VMs, thereby reducing response time and balancing loads across distributed nodes. Additionally, the framework introduces the dynamic optimal load-aware service broker strategy to optimize task scheduling among cloud brokers, aiming to minimize costs, improve energy efficiency, and meet QoS requirements. Similarly, the work~\cite{oudaa2021agent} employs a Quantile Regression-Deep Q Network (QR-DQN) algorithm within a multi-agent system to manage task allocation and resource provisioning, with a focus on reducing energy consumption while maintaining QoS in large-scale data centers. Furthermore, the work~\cite{asghariCombinedUseCoral2022} presents a hybrid approach that combines the the Coral Reef Optimization (CRO) algorithm with a Multi-Agent DQN (MDQ-CR) to enhance energy-aware resource provisioning using DVFS technology in cloud data centers. This two-phase model employs CRO for initial resource allocation and then utilizes a Multi-Agent Deep Q-Network for long-term resource management, addressing high energy demands by avoiding local optima and emphasizing long-term optimization.

As edge computing systems face increasingly complex challenges, MADRL has become a crucial method for addressing resource provisioning in distributed, and heterogeneous environments. For example, the work~\cite{zhangDistributedMultiCloudMultiAccess2021} addresses resource provisioning challenge in a distributed multi-cloud, MEC network by leveraging a MADRL approach. It considers a three-layer architecture involving cloud centers (CCs), MEC servers, and edge devices (EDs), where tasks are distributed among independent CCs that rely on MEC servers and EDs for data processing to minimize latency. The proposed solution optimizes task offloading and resource provisioning in a decentralized manner by allowing each CC to predict and adapt to the resource usage of other CCs, thus reducing system latency and enhancing resource utilization across heterogeneous devices in real time. Additionally, the work~\cite{li2023adaptive} introduces a MADDPG algorithm for resource provisioning in edge network slicing, aiming to balance latency and energy consumption. The proposed framework addresses the limitations of static slicing and instantaneous reward maximization by implementing a novel incremental learning scheme, which adapts the algorithm to dynamic changes in the number of slices without retraining from scratch. Furthermore, the work~\cite{asimdynamic} proposes a MADRL approach for dynamic and efficient resource provisioning in 5G end-to-end network slicing, integrating MEC with network function virtualization (NFV) to support diverse user equipment (UE) demands in densely populated areas like airports and train stations. This strategy aims to maximize service provider profitability, uphold SLAs, and reduce operational costs by effectively managing network slices across distributed cloudlets in MEC environments.

\subsubsection{Advanced DRL for Resource Provisioning}

\begin{table*}[!htbp]
\centering
\caption{Summary of typical works on DRL-based resource scheduling}
\label{tab:literature_resource_scheudling}
    \setlength{\tabcolsep}{3.2mm}{
    \resizebox{\linewidth}{!}{
    \begin{tabular}{cccccccc}
        \toprule
        % \multicolumn{1}{c}{Reference} & \multicolumn{1}{c}{Method} & \multicolumn{1}{c}{Others} & \multicolumn{1}{c}{Environment} & \multicolumn{4}{c}{Optimization objective} \\
        % & & & & Makespan & System Cost & Energy consumption & Others \\
        \multicolumn{1}{c}{\multirow{2}{*}{Reference}} & \multicolumn{1}{c}{\multirow{2}{*}{Method}} & \multicolumn{1}{c}{\multirow{2}{*}{Others}} & \multicolumn{1}{c}{\multirow{2}{*}{Environment}} & \multicolumn{4}{c}{Optimization objective} \\
        & & & & Makespan & Cost & Energy consumption & Others \\
        \midrule
        
        ~\cite{liuHierarchicalFrameworkCloud2017} & DQN & - & Cloud computing & \checkmark & - & \checkmark & - \\
        % ~\cite{cheng2018drl} & DQN & - & cloud computing & - & - & \checkmark & reject rate \\
        ~\cite{kardani-moghaddamADRLHybridAnomalyAware2021} & DQN & - & Cloud computing & \checkmark & - & - & Resource utilization \\
        ~\cite{chenResourceAllocationWorkloadTime2023} & DQN & - & Cloud computing & - & \checkmark & - & QoS \\
        ~\cite{bitsakosDERPDeepReinforcement2018} & DDQN & - & Cloud computing & \checkmark & \checkmark & - & System throughput \\
        ~\cite{zhangComputingResourceAllocation2021} & DQN & - & Edge computing & - & - & - & Overhead and latency \\
        ~\cite{cuiDeepReinforcementLearningBased2023} & DQN & - & Edge computing & \checkmark & - & - & Network latency \\
        ~\cite{liComputingOffloadingResource2021} & DQN & - & Edge computing & - & - & - & Delay and computational cost \\
        ~\cite{fangDeepReinforcementLearningBasedResourceAllocation2022} & DQN & - & Edge computing & - & - & - & Network delay \\
        ~\cite{zhangSecurityComputingResource2024} & DQN & - & Edge computing & \checkmark & \checkmark & - & - \\
        ~\cite{keDeepReinforcementLearningbased2021} & DDQN & - & Edge computing & \checkmark & \checkmark & \checkmark & Bandwidth cost \\
        ~\cite{ullahOptimizingTaskOffloading2023} & DDQN & - & Edge computing & - & - & - & Resource utilization; cost of bandwidth \\
        ~\cite{liuDeepReinforcementLearning2023} & PDQN & - & Edge computing & \checkmark & - & - & Latency \\
        ~\cite{maoResourceManagementDeep2016} & REINFORCE & - & Cloud computing & \checkmark & - & - & - \\
        ~\cite{arvindhanAdaptiveResourceAllocation2023} & Actor-Critic & - & Cloud computing & \checkmark & - & - & Throughput; resource utilization \\
        ~\cite{chenLearningBasedResourceAllocation2019} & A2C & - & Cloud computing & - & - & - & Job latency \\
        ~\cite{wlodzimierzManagementHeterogeneousCloud2021} & PPO & - & Cloud computing & - & \checkmark & - & SLA requirements \\
        % ~\cite{anousheeIntelligentResourceManagement2024} & SARSA and Q-learning & - & edge computing & - & - & \checkmark & utility of service requests and resources \\
        ~\cite{heBlockchainBasedEdgeComputing2021} & A3C & - & Edge-Cloud collaboration & - & - & - & Delay; task drop rate \\
        ~\cite{zhuDeepReinforcementLearningBased2022} & DDPG & - & Edge computing & - & \checkmark & - & Task offloading \\
        ~\cite{huangJointComputationOffloading2023} & TD3 & - & Edge computing & - & \checkmark & - & Task offloading; processing delay \\
        ~\cite{belgacemIntelligentMultiagentReinforcement2022} & MADRL & Q-learning & Cloud computing & - & - & \checkmark & Fault tolerance; workload balancing \\
        ~\cite{narantuyaMultiAgentDeepReinforcement2022} & MADRL & - & Cloud computing & - & - & - & Resource utilization \\
        ~\cite{nagarajanMultiAgentDeep2023} & MADRL & - & Cloud computing & \checkmark & - & \checkmark & Resource utilization \\
        ~\cite{liuMultiagentReinforcementLearning2020} & MADRL & - & Edge computing & - & \checkmark & \checkmark & Computation offloading \\
        ~\cite{rosenbergerDeepReinforcementLearning2022} & MADRL & - & Edge computing & \checkmark & \checkmark & - & Resource usage \\
        ~\cite{keMultiAgentDeepReinforcement2022} & MADRL & DDQN & Edge computing & - & \checkmark & - & Delay; bandwidth \\
        ~\cite{li2023task} & MADRL & - & Edge computing & \checkmark & - & - & - \\
        ~\cite{guoCloudResourceScheduling2021} & REINFORCE & Imitation learning & Cloud computing & \checkmark & - & - & - \\
        ~\cite{zhangIntelligentCloudResource2017} & DQN & Simulated annealing & Cloud computing & - & \checkmark & - & - \\
        ~\cite{weiMultiDimensionalResourceAllocation2023} & DRL variant & NESRL & Cloud computing & - & - & - & Resource utilization balance \\
        ~\cite{jiangStackedAutoencoderBasedDeep2020} & DRL variant & 2r-SAE; ASA; 2pER & Edge computing & \checkmark & - & - & - \\
        ~\cite{xueDeepReinforcementLearning2022} & DQN & GA & Edge computing & \checkmark & - & - & - \\
        ~\cite{aduansereEnergyEfficientOptimizationMobile2024} & QRL & Grover search & Edge computing & - & - & \checkmark & - \\
        ~\cite{silviriantiLayerwiseQuantumDeep2024} & LQ-DRL & - & Edge computing & - & - & \checkmark & QoS \\

        \bottomrule
        \end{tabular}}}
\end{table*}

Advanced DRL techniques have expanded traditional DRL methods by integrating complementary approaches such as QRL~\cite{xuGenerativeAIenabledQuantum2024} and LSTM-augmented DRL~\cite{shahidinejadJointComputationOffloading2020}, offering innovative solutions to complex resource provisioning challenges. In the context of cloud computing, these techniques have enabled the development of sophisticated strategies to enhance resource efficiency. For example, the work~\cite{asghariTaskSchedulingResource2021} introduces a hybrid method that combines parallel SARSA RL agents with GA to improve resource provisioning. This approach focuses on reducing makespan and improving resource utilization within cloud environments. Furthermore, the work~\cite{xuGenerativeAIenabledQuantum2024} presents an intelligent resource provisioning method designed for QRL, particularly focusing on generative AI applications across cloud, edge and mobile nodes. Specifically, this framework utilizes QNNs to represent policies, allowing the RL agents to optimize resource provisioning by taking advantage of quantum-specific phenomena, such as superposition and entanglement. By utilizing QRL, the framework achieves faster convergence rates, reduced decision-making latency, and enhanced stability, effectively improving resource management.

The development of advanced DRL techniques has significantly contributed to addressing the unique challenges posed by distributed architectures in edge computing environments. For instance, the work~\cite{shahidinejadJointComputationOffloading2020} introduces a hybrid method that integrates learning automata (LA), LSTM, and RL to facilitate dynamic resource provisioning decisions. Specifically, LA optimizes action selection by adjusting probabilities based on past outcomes and feedback from the environment, ultimately choosing the action with the highest probability. LSTM models analyze historical data to forecast future request volumes, enhancing the ability of RL agent to make timely and effective resource provisioning decisions. Additionally, the work~\cite{heIntelligentProvisioningVirtualized2022} proposes an intelligent VNF configuration framework for the Cloud of Things. This method enhances the AC model by modifying the the loss function of the agent, which enables stable, monotonic improvements and accelerates convergence, essential for achieving reliable and efficient resource provisioning. To support sustainable resource provisioning for VNFs across multi-network operators, the work~\cite{dieyeDRLBasedGreenResource2023} combines DQN with MCTS. This method filters subcarrier(s)-end-user associations early in the decision process, significantly reducing the computational complexity of DQN for resource provisioning and expediting decision-making. By integrating MCTS, the approach optimizes resource provisioning to meet the computational demands of VNFs in edge networks. Similarly, the work~\cite{chen2019iraf} introduces the iRAF, a resource provisioning model specifically tailored for MEC in IoT environments. By utilizing multitask DRL combined with MCTS, iRAF dynamically allocates resources across edge servers to manage latency and energy requirements in IoT applications. Adapting to real-time changes in network conditions, this framework enhances performance, optimizing resource provisioning in response to complex, fluctuating demands.

\subsection{DRL-Based Resource Scheduling}

Resource scheduling is a critical aspect of optimizing the performance and efficiency of computing systems, encompassing both cloud and edge environments. Subsequently, we provide a comprehensive overview of various DRL-based resource scheduling methods, detailing their specific applications and contributions to resource management within cloud and edge computing settings. These studies are systematically categorized in Table~\ref{tab:literature_resource_scheudling}, offering an in-depth summary of recent advancements in this field. 

% Given the escalating demand for computational resources, the design and implementation of effective resource scheduling mechanisms are imperative for ensuring dynamic and efficient resource management.

\subsubsection{Value-Based DRL for Resource Scheduling}

Value-based DRL methods concentrate on assessing the value of actions to optimize decision-making, providing a powerful approach for tackling resource scheduling challenges in dynamic and distributed computing environments~\cite{liuHierarchicalFrameworkCloud2017, kardani-moghaddamADRLHybridAnomalyAware2021, bitsakosDERPDeepReinforcement2018}.
DQN, as one of the foundational methods in value-based DRL, has been widely utilized to address resource scheduling problems in cloud computing environments. For instance, the work~\cite{liuHierarchicalFrameworkCloud2017} develops a joint VM resource scheduling and power management framework for cloud computing systems, utilizing DQN to efficiently handle high-dimensional state and action spaces. The framework aims to minimize power consumption while ensuring acceptable performance levels. Furthermore, the work~\cite{kardani-moghaddamADRLHybridAnomalyAware2021} proposed a hybrid anomaly-aware DQN-based resource scaling method for dynamic resource allocation in cloud environments. DQN is employed to ensure efficient resource utilization and minimize makespan, while the approach also integrates anomaly detection to enhance decision-making stability in response to anomalous states. Additionally, the work~\cite{chenResourceAllocationWorkloadTime2023} introduces a resource scheduling approach for cloud-based software services. This method employs a DQN-based prediction model trained using workload-time windows to anticipate management operations, specifically adjusting the number of VMs in response to varying system states. However, the inherent overestimation bias in traditional DQN can hinder learning stability. To address this, the work~\cite{bitsakosDERPDeepReinforcement2018} incorporates DDQN into the proposed Deep Elastic Resource Provisioning (DERP) method. This approach aims to enhance elasticity in large-scale computing clusters by optimizing metrics such as makespan, resource cost, and system throughput.

Beyond cloud computing, value-based DRL approaches have also demonstrated its utility in edge computing resource scheduling. For example, the work~\cite{zhangComputingResourceAllocation2021} addresses the complexities of computing task offloading in MEC for the Internet of Vehicles (IoV) by developing a DQN-based resource scheduling scheme. The framework considers service node computational capabilities and vehicle velocity to achieve minimized system overhead and latency. Likewise, the work~\cite{cuiDeepReinforcementLearningBased2023} designs a DQN-driven, cloud-edge cooperative strategy for content delivery within the IoV framework, effectively reducing network latency by optimizing caching and routing decisions. This approach leverages historical request patterns and real-time network states, enabling adaptive content management across network nodes. Furthermore, the work~\cite{liComputingOffloadingResource2021} introduces a DQN-based computing offloading resource scheduling strategy to address challenges such as increased latency, energy consumption, and reduced service quality in vehicular networks. Moreover, the work~\cite{fangDeepReinforcementLearningBasedResourceAllocation2022} proposes a DQN-based resource scheduling scheme to enhance content distribution in a layered fog radio access network (FRAN). It addresses the challenge of low-latency content transmission by formulating an optimal resource allocation problem as a minimal delay model, utilizing in-network caching to aggregate content requests. Furthermore, the work~\cite{zhangSecurityComputingResource2024} proposes an action-constrained DQN approach for secure computing resource scheduling in serverless cloud-edge computing environments, with the objective of reducing overall system costs and makespan, while maintaining security in resource utilization.

Given the limitations of traditional DQN in managing the complexities of edge computing, several DQN variants have been proposed to improve performance in these scenarios. For instance, the work~\cite{keDeepReinforcementLearningbased2021} employs DDQN to address the challenge of insufficient processing capabilities in wireless devices. By introducing a model for partial computation offloading and resource scheduling in mobile edge computing, this approach aims to optimize the system's weighted sum cost, computation delays, power consumption, and bandwidth utilization. 
Likewise, the work~\cite{ullahOptimizingTaskOffloading2023} proposes an edge-cloud resource scheduling framework leveraging DDQN to dynamically assess both resource utilization and network conditions, thereby enabling optimal offloading decisions that minimize delays and fulfill computational and communication requirements. Another work~\cite{liuDeepReinforcementLearning2023} addresses the unique challenges of computation-intensive task processing on edge devices by proposing a method based on Parameterized DQN (PDQN) for service placement and resource scheduling. In this approach, service placement involves discrete decisions, whereas resource scheduling demands continuous decisions. PDQN handles this mixed action space by defining a deterministic function that maps states to the continuous parameters associated with each discrete action.

\subsubsection{Policy-Based DRL for Resource Scheduling}

Policy-based DRL methods directly learn the policy that maps states to actions, demonstrating significant advantages in optimizing resource scheduling across diverse environments. Techniques such as REINFORCE~\cite{maoResourceManagementDeep2016}, AC~\cite{arvindhanAdaptiveResourceAllocation2023}, A2C~\cite{chenLearningBasedResourceAllocation2019} and PPO~\cite{funika2020management} have been explored in cloud computing for their efficiency and adaptability. For instance, the work~\cite{maoResourceManagementDeep2016} introduces the DeepRM method, which represents the scheduling policy through a DNN and trains it using the REINFORCE algorithm to minimize average job slowdown. Furthermore, the work~\cite{arvindhanAdaptiveResourceAllocation2023} proposes an adaptive resource scheduling method based on AC to address the load balancing optimization problem in cloud data centers. According to the strategy learned by the actor, appropriate VMs are allocated to tasks, thereby optimizing makespan, response time, resource utilization, and throughput. Compared to REINFORCE, the AC method reduces variance, leading to faster convergence and improved stability. Moreover, the work~\cite{chenLearningBasedResourceAllocation2019} applies the A2C method to resource scheduling in data centers, dynamically adjusting task assignments based on current resource utilization to minimize task latency. By incorporating the advantage function, A2C reduces variance in policy gradient estimates, resulting in more stable learning and less noisy updates. PPO has also been applied to enhance sample efficiency in resource scheduling. The work~\cite{funika2020management} presents a novel PPO-based approach for automated resource scheudling in heterogeneous cloud environments. This method enables the PPO agent to interact with a data center environment containing large, medium, and small VMs, learning policies to maximize resource cost-effectiveness and meet SLA targets.

As edge computing gains prominence, policy-based DRL has also been adapted to meet the distinct demands of edge and IoT networks. For instance, in the field of blockchain, the work~\cite{heBlockchainBasedEdgeComputing2021} addresses security challenges in edge-centric computing by utilizing a blockchain-based framework that integrates the A3C algorithm for resource scheduling. This approach enhances trust in the system while minimizing delays and task drop rates, providing a secure and efficient solution for resource scheduling in decentralized networks. In the realm of VEC, the work~\cite{zhuDeepReinforcementLearningBased2022} formulates a two-stage Stackelberg game to incentivize computation offloading by VEC servers, using a DDPG-based resource scheduling scheme. The DDPG algorithm effectively manages continuous action spaces, allowing both vehicles and servers to optimize their resource utilization and maximize profits. Furthermore, the work~\cite{huangJointComputationOffloading2023} proposes a joint computation offloading and resource scheduling framework in IoV. This method leverages the TD3 algorithm, which offers improved stability and faster convergence in handling continuous action spaces, thereby reducing system costs while maintaining high performance.

\subsubsection{Multi-Agent DRL for Resource Scheduling}

% MADRL enables multiple agents to learn and interact within a shared environment, making it particularly suitable for addressing the complexities of distributed systems. By allowing agents to coordinate and adapt to resource competition, MADRL facilitates effective decentralized decision-making and collaborative strategies, which are essential in dynamic and resource-intensive scenarios. 
MADRL enables multiple agents to learn and interact within a shared environment, making it particularly effective for tackling the complexities of distributed systems. Its application in cloud computing has opened new avenues for improving efficiency and adaptability in resource scheduling. For example, the work~\cite{belgacemIntelligentMultiagentReinforcement2022} combines multi-agent framework with Q-learning by treating VMs as agents that dynamically adjust their states to meet requirements for energy consumption, fault tolerance, and load balancing, thereby optimizing overall system performance. In a similar vein, the work~\cite{narantuyaMultiAgentDeepReinforcement2022} explores the configuration of a high-performance AI computing environment using advanced technologies, such as Nvidia DGX servers, to leverage MADRL for regional resource utilization optimization, resulting in more effective resource scheduling. Additionally, the work~\cite{nagarajanMultiAgentDeep2023} focuses on energy efficiency in cloud environments by utilizing container-based virtualization, which offers higher efficiency than traditional VMs. This approach employs MADRL to dynamically schedule tasks and allocate resources, significantly reducing energy consumption while accounting for VM overheads and workload patterns.

As edge computing and IoT networks grow in complexity, MADRL has become a vital tool for optimizing resource scheduling in these domains. For instance, the work~\cite{liuMultiagentReinforcementLearning2020} addresses resource competition in IoT edge computing by treating users as independent learning agents within a dynamic and uncertain environment. This approach uses a multi-agent Q-learning algorithm, allowing users to make optimal offloading decisions independently, minimizing long-term system costs without needing to know the actions of others. Similarly, another work~\cite{rosenbergerDeepReinforcementLearning2022} applies DRL to industrial IoT (IIoT) systems, employing multi-agent systems for decentralized decision-making, where one agent is responsible for device resource scheduling and the other manages network resource scheduling. This dual-agent design enhances scalability and adaptability to changing numbers of network nodes, providing greater flexibility in resource scheduling. In the context of 5G networks, the work~\cite{keMultiAgentDeepReinforcement2022} proposes a MEC framework utilizing a decentralized MADRL algorithm to address the challenges of low latency and high reliability in resource scheduling. By considering task priorities and channel variability, this framework effectively minimizes long-term costs related to delay and bandwidth, allowing edge clouds to support machine learning tasks with optimized resource scheduling. Furthermore, the work~\cite{li2023task} introduces a distributed scheduler for resource allocation in edge computing networks, leveraging a GNN-based MADRL paradigm to address the complexities of resource scheduling for machine learning tasks in edge environments. This framework includes a heterogeneous graph attention network to manage interactions among distributed agents, along with a task selection mechanism and conflict resolution strategies, enhancing scheduling performance in multi-task scenarios across distributed edge clusters. By aiming to minimize task completion time and improve resource utilization efficiency, this approach offers significant advancements in edge resource management.

\subsubsection{Advanced DRL for Resource Scheduling}

Advanced DRL methods have increasingly integrated diverse techniques to enhance learning efficiency and address complex resource scheduling challenges. By incorporating approaches such as imitation learning~\cite{guoCloudResourceScheduling2021}, evolutionary strategies~\cite{zhangIntelligentCloudResource2017}, or quantum computing~\cite{silviriantiLayerwiseQuantumDeep2024}, these methods tackle sophisticated resource scheduling challenges, achieving superior optimization and adaptability. In cloud computing, several studies have explored such techniques. For instance, the work~\cite{guoCloudResourceScheduling2021} integrates imitation learning with the REINFORCE algorithm to accelerate model training and convergence. Imitation learning mimics the behavior of heuristic algorithms, allowing the RL agent to leverage expert strategies from these heuristics instead of relying on random exploration, significantly speeding up the learning process. Alternatively, some studies have combined heuristic algorithms with DRL. For example, the work~\cite{zhangIntelligentCloudResource2017} addresses cloud service configuration and adaptive resource scheduling challenges by integrating a modified DQN algorithm with SA. SA enhances DRL by gradually lowering the exploration temperature, allowing the agent to explore a broader range of actions in the early training stages and then prioritize exploitation as training progresses. This temperature-controlled exploration mechanism enables the agent efficiently balance exploration and exploitation, thereby improving convergence speed and enabling more effective resource scheduling in dynamic cloud environments. Additionally, the work~\cite{weiMultiDimensionalResourceAllocation2023} introduces an advanced DRL-based method that integrating Natural Evolution Strategy (NES) into A3C to optimize resource scheduling across distributed data centers. By utilizing NES to approximate the gradient of the reward function, this method enhances training efficiency and maintains exploration diversity, resulting in a more balanced utilization of resources. This approach effectively addresses the challenges of multi-dimensional resource allocation, achieving improved performance in distributed cloud environments.

Some studies have proposed advanced DRL methods specifically for IoT and edge computing environments to address resource scheduling challenges. For instance, the work~\cite{jiangStackedAutoencoderBasedDeep2020} introduces an advanced RL framework for online resource scheduling in large-scale MEC systems. This is enhanced by a related and regularized stacked auto-encoder (2r-SAE) for data compression, an adaptive SA approach for search efficiency, and a preserved and prioritized experience replay (2pER) mechanism to improve policy training. RL can also assist heuristic algorithms, for example, the work~\cite{xueDeepReinforcementLearning2022} combines DQN with GA to improve resource scheduling on edge cloud servers, minimize application execution time. In this approach, DQN generates the initial population for GA, reducing computational costs and improving optimization for the scheduling problem. Some studies further leverage QRL to boost resource scheduling performance. For example, the work~\cite{aduansereEnergyEfficientOptimizationMobile2024} proposes an innovative QRL method for dynamic IoT networks that jointly optimizes resource scheduling, content caching, and computation offloading to maximize energy efficiency. Leveraging quantum computing techniques, including an improved Grover search algorithm, this method accelerates the training process and increases policy adaptability within high-dimensional continuous action spaces, thereby significantly boosting overall effectiveness. Similarly, the work~\cite{silviriantiLayerwiseQuantumDeep2024} develops a layerwise QRL to address continuous large-space and time-series challenges for resource scheduling. By utilizing quantum embeddings, this framework focuses on UAV trajectory planning and power allocation, aiming to optimize energy consumption while ensuring QoS. 

%------------------------------------------------------------------------
\section{Future Development Trends}
\label{sec:trends}
%------------------------------------------------------------------------
Building on the review of existing works in DRL-based job scheduling and resource management, this section discusses the challenges of applying DRL in cloud computing and provides insights into potential future directions.

\subsection{Privacy and Security for Job Scheduling}\label{4.1}
As cloud computing continues to expand, job scheduling faces significant challenges in addressing privacy and security concerns, particularly in scenarios involving sensitive data processing and cross-platform collaboration. Privacy protection, especially in shared cloud environments, requires the implementation of strict data and task privacy constraints during scheduling. To address this, one promising direction involves leveraging hybrid cloud architectures, which enable resource isolation and elastic management to support dynamic partitioning and cross-data-center scheduling~\cite{sun2023efficient, sun2022et2fa}. Another critical avenue is the integration of privacy protection directly into the objective function of scheduling algorithms, allowing for a balanced optimization of privacy, performance, and cost.~\cite{wen2020scheduling}  Security remains a parallel concern, with the confidentiality and integrity of sensitive data posing significant challenges. Current practices, such as employing encryption algorithms like RC4 for confidentiality~\cite{alenezi2020symmetric} and SHA-1 for integrity~\cite{prasanna2021performance}, offer foundational safeguards for intermediate and stored data. Future research will focus on refining these security measures, particularly by integrating them into the scheduling process. This includes addressing issues such as user authentication, secure resource allocation, and scalable security protocols to ensure robust protection for sensitive data in increasingly complex and heterogeneous cloud environments.

\subsection{Resource Management over Multi-Tier Networks}
Modern computing environments increasingly rely on multi-tier networks to address growing complexities and diverse application demands. These networks, comprising edge, fog, mist, and dew computing layers, pose significant challenges for resource management due to their heterogeneous resource characteristics and varied latency requirements. Addressing these challenges requires advanced frameworks and innovative technologies tailored to the unique needs of multi-tier architectures. One promising direction is the integration of DT technology into resource management~\cite{zhou2022digital}. DT technology provides synchronized virtual representations of physical entities, enabling real-time monitoring, predictive analysis, and proactive management of resource utilization and user demands. Another significant future trend is the adoption of multi-tier multi-domain network slicing~\cite{oladejo2021multi, ou2023two}. This approach facilitates resource aggregation across multiple infrastructure providers, allowing for more effective distribution of resources across domains and the implementation of tailored strategies that meet the specific requirements of diverse applications.

% The future of resource management is increasingly focused on multi-tier networks due to their ability to address the complexities of modern computing environments. Comprising edge, fog, mist, and dew computing layers, multi-tier networks necessitate sophisticated strategies to effectively manage the heterogeneous nature of resources while accommodating the diverse latency requirements of various applications. One prominent trend in this area is the integration of digital twin technology into resource management frameworks~\cite{zhou2022digital}. Digital twins provide synchronized virtual representations of physical entities, facilitating real-time monitoring and predictive analysis of resource utilization and user requests. Another significant future trend is the adoption of multi-tier multi-domain network slicing for resource management~\cite{oladejo2021multi, ou2023two}. This approach enables the aggregation of resources from multiple infrastructure providers, facilitating more effective resource distribution across domains and allowing for tailored resource management strategies that meet the distinct requirements of various applications.

\subsection{Handling Large-Scale Decision-Making in Job Scheduling and Resource Management}
In large-scale job scheduling and resource management scenarios, such as multi-cloud data centers, the inherent complexity of these environments introduces significant challenges. For single-agent systems, the primary obstacle lies in the overwhelming state and action spaces, commonly referred to as the ``curse of dimensionality". The high dimensionality of these spaces increases the computational complexity and hampers the learning efficiency of DRL algorithms. Hierarchical reinforcement learning emerges as a promising solution to address these challenges by decomposing job scheduling or resource management into multiple decision-making layers~\cite{guan2022hierrl, zhang2024adaptive}. High-level agents manage coarse-grained decisions, while low-level agents handle finer-grained decisions. This layered approach simplifies individual agent tasks, improving both learning efficiency and scalability. In multi-agent systems, coordinating hundreds or thousands of agents introduces further challenges. Traditional centralized approaches face significant limitations, including high communication overhead, and latency, alongside scalability issues as system size increases. To overcome these limitations, decentralized model-based policy optimization frameworks have emerged as a promising research direction~\cite{ma2024efficient}. These frameworks enable agents to interact primarily with immediate neighbors, significantly reducing communication costs while maintaining effective coordination. By leveraging localized interactions and model-based learning, decentralized approaches enhance scalability and efficiency. This paradigm allows agents to achieve robust decision-making without relying on extensive global communication, making it particularly suitable for large-scale and complex environments.

% Many existing job scheduling approaches characterize the action space in a manner where its dimensionality directly corresponds to the number of available VMs. In large-scale infrastructures, such as multi-cloud data centers, this leads to an excessively large action space, resulting in the ``curse of dimensionality", which significantly impairs the learning efficiency of DRL algorithms. Thus, formulating an efficient action space is crucial for enhancing DRL performance in job scheduling. Hierarchical reinforcement learning (HRL) presents a promising solution by decomposing the scheduling task into multiple decision-making layers~\cite{guan2022hierrl, zhang2024adaptive}. Within this hierarchical architecture, high-level agents focus on coarse-grained decisions, such as assigning workloads to specific availability zones or data centers, while low-level agents manage finer-grained decisions, including selecting individual nodes or VMs for task execution. This approach effectively reduces the dimensionality of the action space and facilitates more efficient learning. Additionally, action branching architectures have proven effective in managing large action spaces~\cite{he2022ddpg}. This technique decomposes the high-dimensional action space into multiple low-dimensional branches, allowing each branch to make decisions independently. By discretizing multi-dimensional continuous actions, the action branching method prevents the exponential growth of action dimensions. 

\subsection{Incorporating Large Language Models (LLMs) for Job Scheduling and Resource Management}% predetermined user request patterns -> heterogeneous edge computing architecture
 
Job scheduling and resource management in dynamic environments face notable challenges due to their complexity and variability. Traditional DRL approaches, while effective in well-defined scenarios, often struggle to adapt to unforeseen situations, relying heavily on training tailored to specific environments, such as fixed-scale resource pools or predetermined user request patterns. This lack of flexibility limits their applicability in diverse and dynamic contexts, while frequent retraining to accommodate changing conditions adds significant computational overhead. To address these challenges, the integration of LLMs represents a transformative research direction. Unlike traditional DRL methods, LLMs possess a remarkable ability to generalize across diverse scenarios, enabling them to adapt to previously unseen environments without the need for exhaustive retraining~\cite{lai2023large}. This adaptability is especially valuable in dynamic systems, where LLMs can interpret changing workloads and resource states in real time, providing more flexible and efficient solutions. Furthermore, their capability for knowledge transfer across different contexts accelerates learning process, reduces deployment overhead, and supports robust decision-making in heterogeneous environments. By leveraging these strengths, LLMs open new avenues for advancing job scheduling and resource management in complex, ever-changing settings.

\section{Conclusion}
\label{sec:con}
This paper presents a comprehensive analysis of advancements in deep reinforcement learning (DRL) methodologies for job scheduling and resource management in cloud computing, with a focus on reviewing existing works categorized by the specific DRL algorithms used. We began by outlining the modeling of these optimization problems using Markov Decision Processes (MDPs) and demonstrated how DRL can effectively address them. Subsequently, we provided an overview of current DRL algorithms and systematically reviewed their applications in job scheduling, including task scheduling and workflow scheduling, as well as resource management, focusing on resource provisioning and scheduling. The reviewed works are categorized based on the DRL algorithms employed, offering a clear framework for understanding their implementation and impact.

In addition to analyzing existing methodologies, we provide insights to guide future research and practical advancements, paving the way for more efficient and adaptive DRL-driven solutions. This review underscores the pivotal role of DRL in tackling the complex and dynamic challenges of job scheduling and resource management. As cloud computing environments continue to evolve, we expect that DRL-based approaches will become increasingly proficient at handling the unpredictable demands of workload and resource allocation, driving significant advancements in these domains.

%------------------------------------------------------------------------
%\section*{Acknowledgments}
%------------------------------------------------------------------------
%The authors would like to thank

%------------------------------------------------------------------------
\bibliographystyle{IEEEtran}
\bibliography{IEEEabrv,survey}
%------------------------------------------------------------------------

%\begin{comment}

\begin{IEEEbiography}
[{\includegraphics[width=1in,height=1.25in,clip,keepaspectratio]{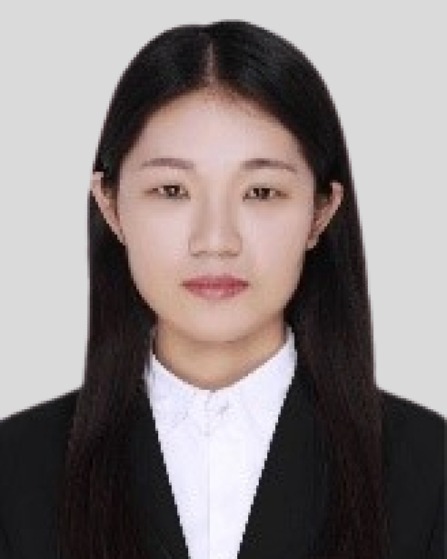}}]
{Yan Gu} received the B.E. degree from Nanjing Institute of Technology, Nanjing, China, in 2020, and M.S. degree from the School of Computer at Jiangsu University of Science and Technology, China. She is currently a PhD student in the School of Control and Computer Engineering at North China Electric Power University in Beijing. Her research interests include cloud computing, deep learning, parallel and distributed computing.
\end{IEEEbiography}

\begin{IEEEbiography}
[{\includegraphics[width=1in,height=1.25in,clip,keepaspectratio]{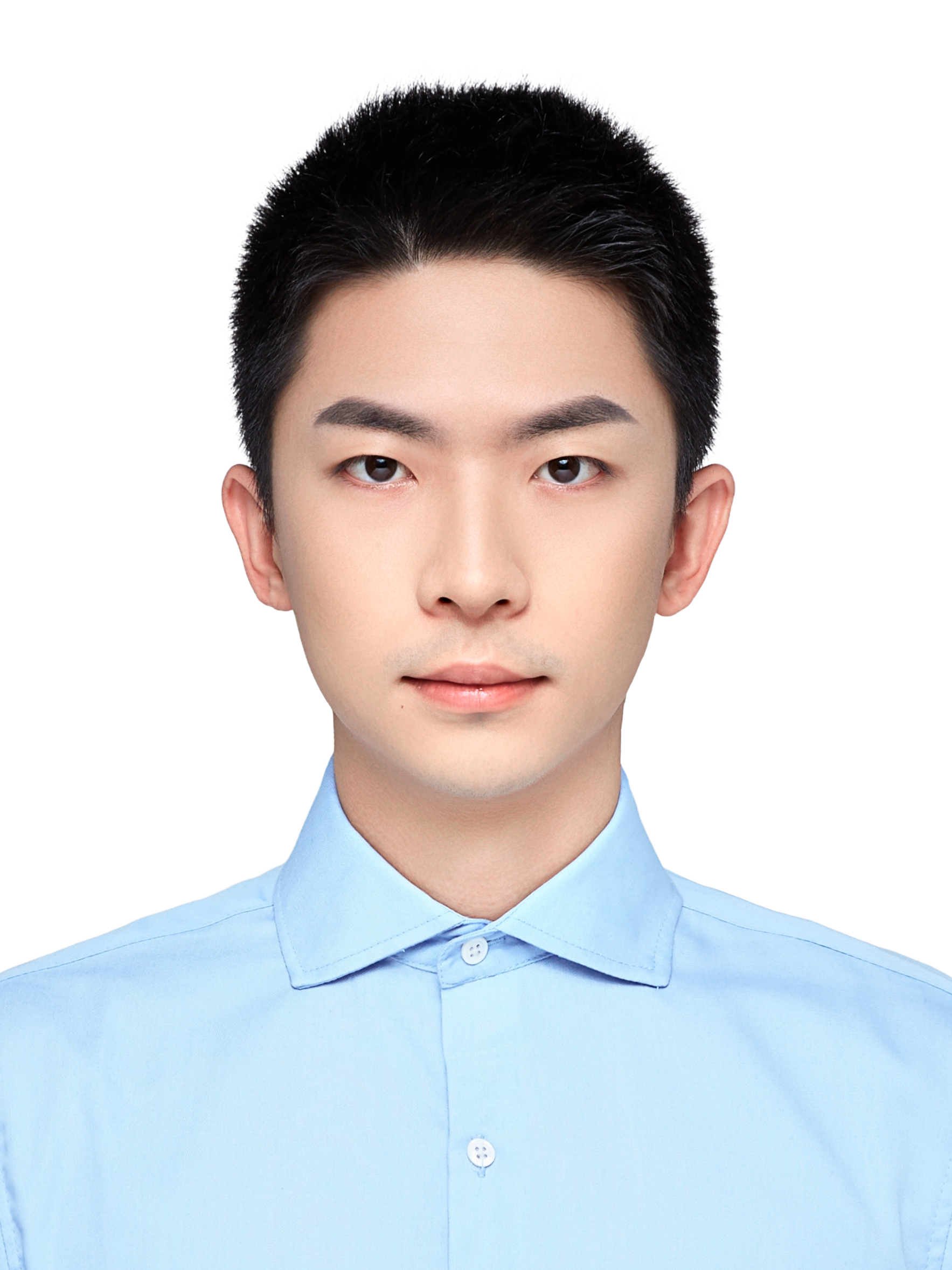}}]{Zhaoze Liu} received the B.E. degree from Beijing University of Posts and Telecommunications, Beijing, China, in 2022. He is currently pursuing the M.S. degree in the School of Control and Computer Engineering at North China Electric Power University in Beijing. His research interests include cloud computing, deep reinforcement learning, and serverless computing.
\end{IEEEbiography}

\begin{IEEEbiography}
[{\includegraphics[width=1in,height=1.25in,clip,keepaspectratio]{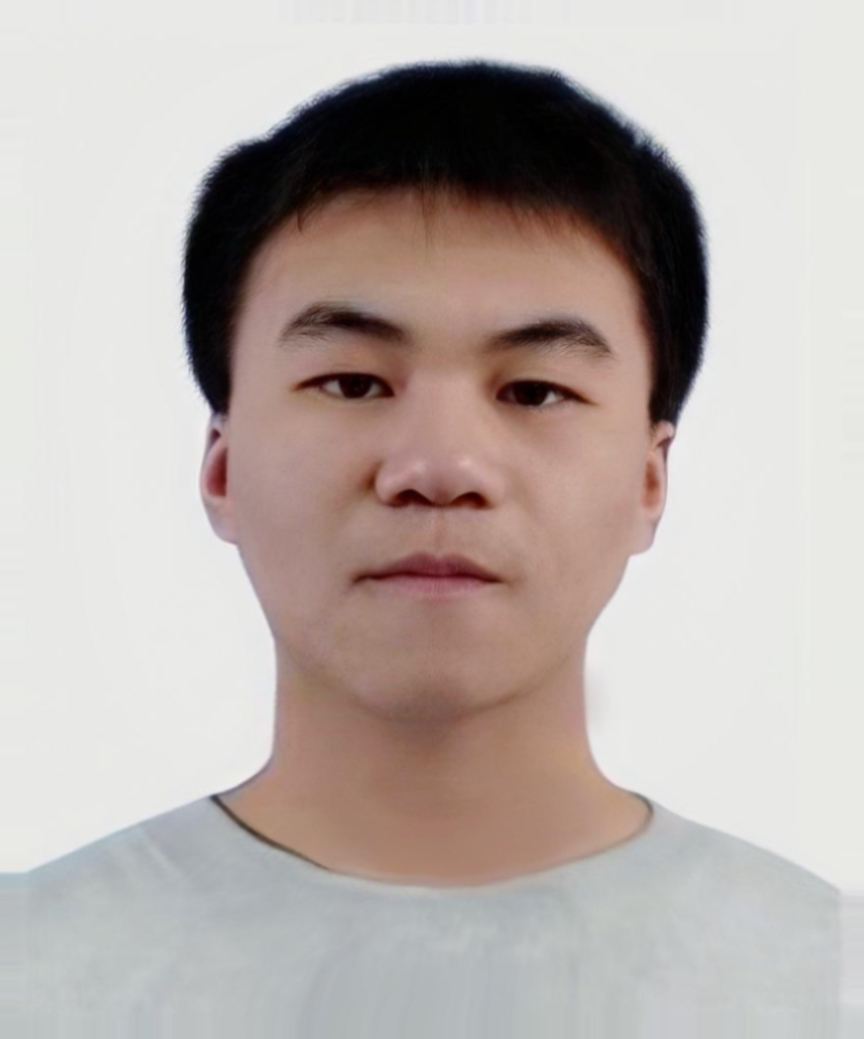}}]{Shuhong Dai} is a master student in Software Engineering at North China Electric Power University in Beijing. He received his B.E. degree in Electrical Engineering and Automation from Ningbo University of Technology, China, in 2023. His research interests include deep reinforcement learning, cloud computing and intelligent transportation systems.
\end{IEEEbiography}

\begin{IEEEbiography}[{\includegraphics[width=1in,height=1.25in,clip,keepaspectratio]{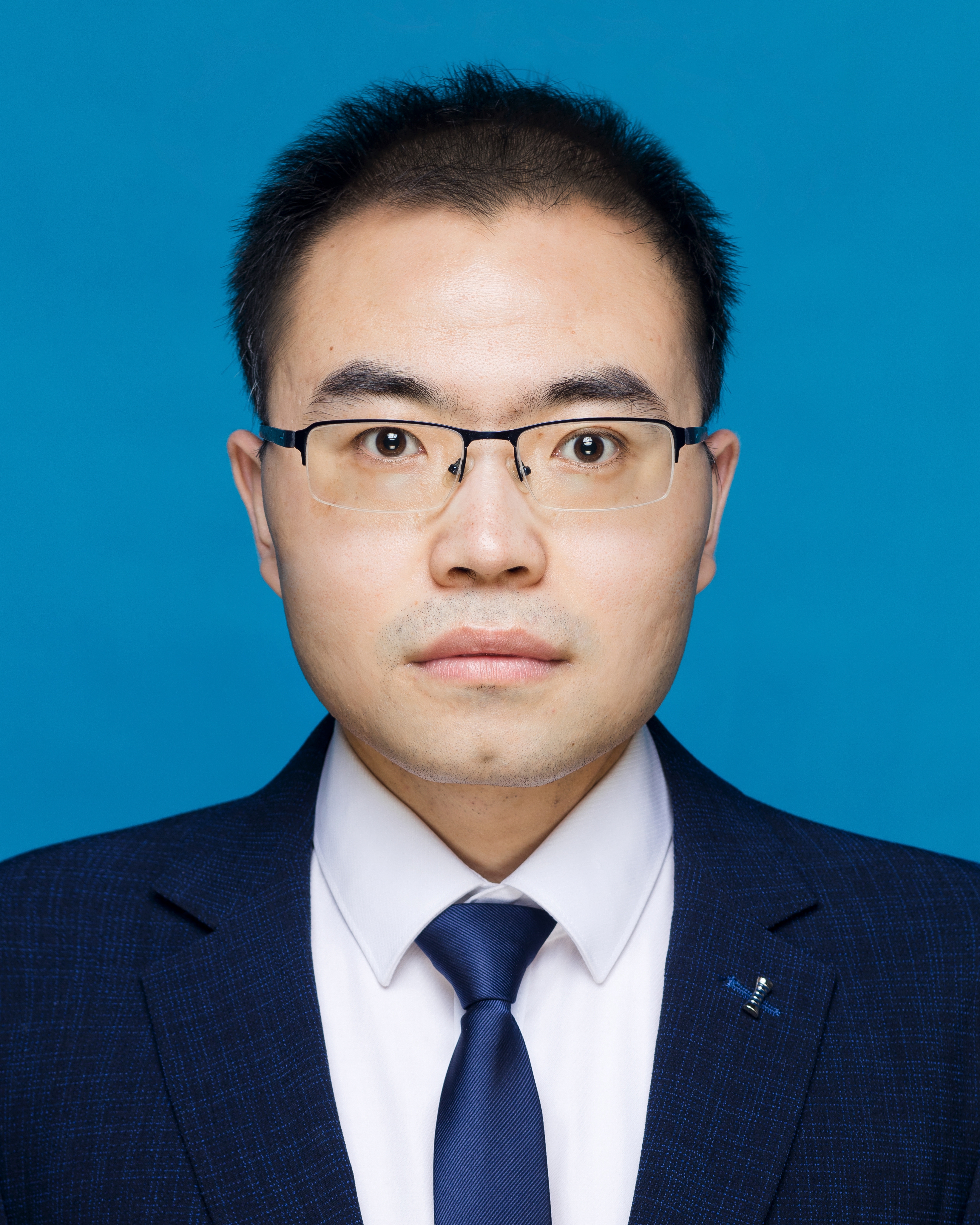}}]{Cong Liu}
received the PhD degree in the Department of Mathematics and Computer Science, Eindhoven University of Technology in 2019. He is a full Professor in Shandong University of Technology. His research interests are in the areas of business process management, process mining, and workflow management.
\end{IEEEbiography}

\begin{IEEEbiography}[{\includegraphics[width=1in,height=1.25in,clip,keepaspectratio]{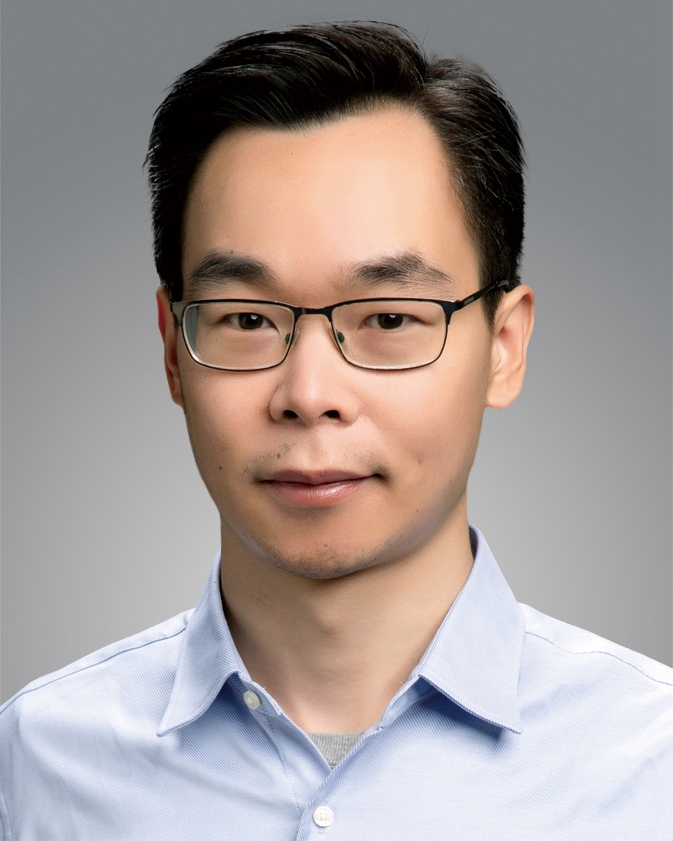}}]{Ying Wang} is a Professor at Institute of Computing Technology (ICT), Chinese Academy of Sciences (CAS). He received Ph.D degree from ICT in 2014. His research interests includes computer architecture and VLSI design, specifically memory system, on-chip interconnects, resilient and energy-efficient architecture, machine learning accelerators, and parallel data systems. 
\end{IEEEbiography}

\begin{IEEEbiography}[{\includegraphics[width=1in,height=1.25in,clip,keepaspectratio]{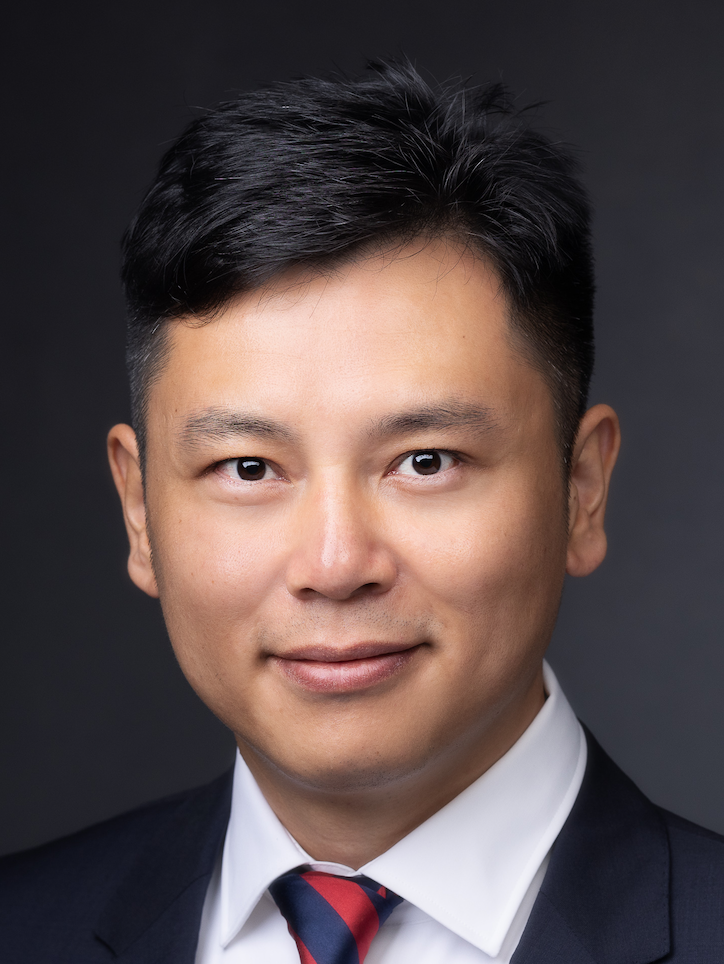}}]{Shen Wang}
is an Assistant Professor with the School of Computer Science, University College Dublin, Ireland. He received the Ph.D. degree from Dublin City University, Ireland. Dr. Wang has been involved with several EU projects as a co-PI, WP and Task leader in big trajectory data streaming for air traffic control and trustworthy AI for intelligent cybersecurity systems. His research interests include connected autonomous vehicles, deep reinforcement learning, and security and privacy for mobile networks.
\end{IEEEbiography}

\begin{IEEEbiography}[{\includegraphics[width=1in,height=1.25in,clip,keepaspectratio]{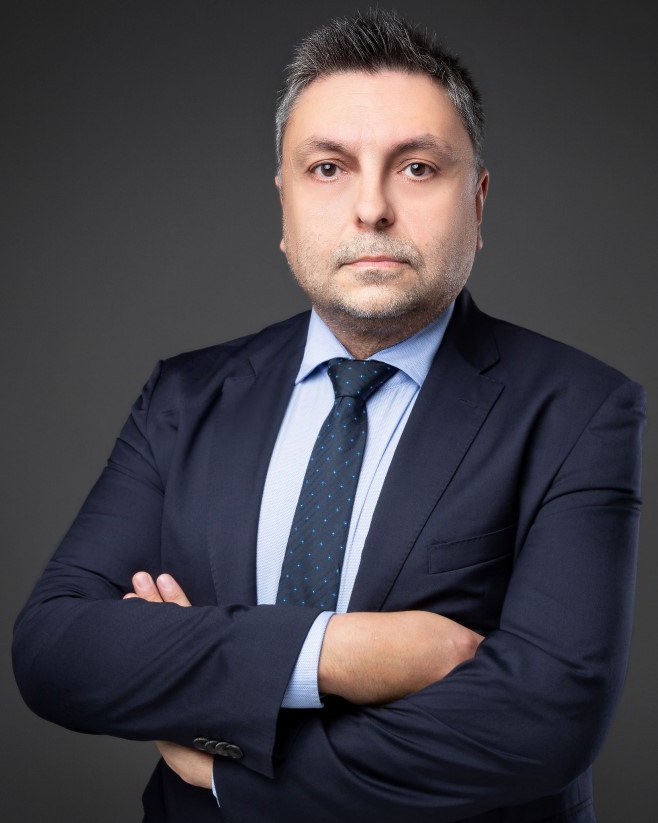}}]{Georgios Theodoropoulos} is currently a Chair Professor at the Department of Computer Science and Engineering at SUSTech in Shenzhen, China. He was previously the inaugural Executive Director of the Institute of Advanced Research Computing and a Chair Professor at the School of Engineering and Computing Sciences at the University of Durham, UK. He has been a Senior Research Scientist with IBM Research and senior faculty at the University of Birmingham, UK, where he was also founding Director of one of UK’s e-Science Centres of Excellence. He has held an Adjunct Chair at the Trinity College Dublin and visiting appointments at the Nanyang Technological University and National University in Singapore.  He is Ordinary Member of the European Academy of Sciences and Arts, a Fellow of the World Academy of Art and Science, an Accredited Board Director of the Singapore Institute of Directors, a Chartered Engineer, and holds a PhD from the University of Manchester, UK.
\end{IEEEbiography}

\begin{IEEEbiography}[{\includegraphics[width=1in,height=1.25in,clip,keepaspectratio]{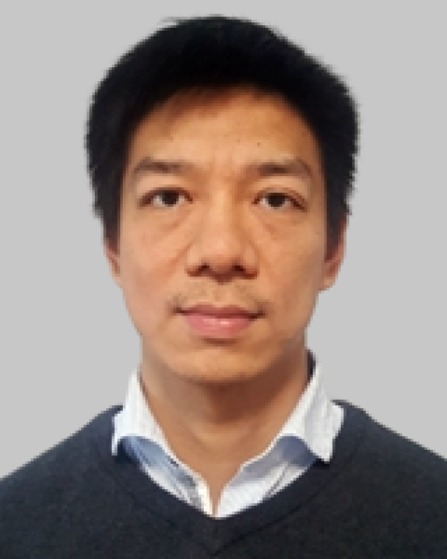}}]{Long Cheng} is a Professor at North China Electric Power University in Beijing.  He received the Ph.D from National University of Ireland Maynooth in 2014. He was an Assistant Professor at Dublin City University, and a Marie Curie Fellow at University College Dublin. He has published more than 110 papers in refereed journals and conferences. His research focuses on distributed computing and deep reinforcement learning. Prof Cheng is an Associate Editor of IEEE Transactions on Consumer Electronics, and a Chair of Journal of Cloud Computing. More info: \url{https://longcheng.eu/}
\end{IEEEbiography}

%\end{comment}

\vfill
\end{CJK}
\end{document}